\documentclass[10pt]{article}

\usepackage[version=3]{mhchem} 

\usepackage{caption}
\usepackage{subcaption}

\usepackage[boxed,commentsnumbered]{algorithm2e} 
\usepackage{comment}
\usepackage{amsmath}
\usepackage{amssymb}
\usepackage{epsfig}
\usepackage{siunitx}
\usepackage{multirow}
\usepackage{url}
\usepackage{fullpage}

\newcommand{\Gridfig}{fig/Grid}
\newcommand{\MCfig}{fig/MC}

\newcommand{\EasalfigJac}{fig/Easal_jacobian}
\newcommand{\EasalfigA}{fig/Easal_differentStepSizeAmongNodes}
\newcommand{\EasalfigW}{fig/Easal_differentStepSizeWithinNode}
\newcommand{\EasalfigWR}{fig/Easal_differentStepSizeWithinNode_Reverse}

\newcommand{\ldire}{fig/extra}
\newcommand{\ldirr}{fig/mc_figs}

\newcommand{\Cp}{Cayley point}

\newcommand{\comm}[1]{}

\newcommand{\figref}[1]{Fig.~\ref{#1}}

\usepackage{color}
\newcommand{\eat}[1]{}

\author{Aysegul Ozkan \and Jose C. Flores-Canales \and Rahul Prabhu \and Meera Sitharam \and Maria Kurnikova}

\title{\Large\bf Fast and Flexible Geometric Method For Enhancing MC Sampling of Compact Configurations For Protein Docking Problem}

\begin{document}

\maketitle

\begin{abstract}
EASAL (Efficient Atlasing and Sampling of Assembly Landscapes) is a geometric
method for sampling and computing integrals over the potential energy landscape
of small molecular assemblies. EASAL's efficiency arises from the fact that
small assembly landscapes permit the use of so-called Cayley 
(inter-atomic distance based) parameters for geometric representation and 
sampling of the assembly configuration space regions; this results in their 
isolation, convexification, customized sampling  and systematic traversal 
using a comprehensive topological roadmap. 
   
We define custom-designed measurements to investigate and compare various
sampling characteristics of EASAL and the traditional Monte Carlo (MC)
sampling, including (i) sampling speed, (ii) efficiency and accuracy of
uniform grid coverage, (iii) accuracy of weighted coverage at covering
low energy regions, (iv) ability to localize sampling to macrostates,
and (v) flexibility in sampling distributions. 

In particular, we compare the sampling characteristics of EASAL and MC in 
sampling the assembly
landscape of 2 trans-membrane helices, with short-range pair-potentials. We
demonstrate that EASAL provides a reasonable coverage of crucial but narrow
regions of the energy landscape of low effective dimension, with much fewer
samples and computational resources than MC sampling. Promising avenues for
combining the complementary advantages of the two methods are discussed.

\eat{Additionally, since accurate computation of configurational entropy and other
integrals is required for estimation of both free energy and kinetics, it is
essential to obtain \emph{uniform} sampling in appropriate Cartesian or moduli
space parameterization.  EASAL's  flexibility is demonstrated with a variety of
sampling distributions, from Cayley sampling skewed towards lower energy
regions, to uniform Cartesian sampling at the two ends of the spectrum. }
    
\end{abstract}

\section{Introduction}
\label{sec:mc_intro}

To overcome the problem of incomplete sampling of relevant phase space when
modeling protein assembly, in this paper, we use the geometric method EASAL
(Efficient Atlasing of Assembly Landscapes)\cite{Ozkan2011, maineasal,
Ozkan2014Jacobian, Ozkan:toms}. In addition, we define custom-designed
measurements to investigate and compare various sampling characteristics of
EASAL and the traditional Monte Carlo (MC) sampling, including (i) sampling
speed, (ii) efficiency and accuracy of uniform grid coverage, (iii) accuracy of
weighted coverage at covering low energy regions, (iv) ability to localize
sampling to macrostates, and (v) flexibility in sampling distributions. 

In particular, we sample the energy landscape of two transmembrane helices
assembling via short-range pair-potentials demonstrate that EASAL provides a
reasonable coverage of crucial but narrow regions of the energy landscape of
low effective dimension, with much fewer samples and computational resources
than MC sampling. Promising avenues for combining the complementary advantages
of the two methods are discussed.

EASAL partitions the energy landscape of transmembrane helices into nearly
equipotential energy regions called \emph{active constraint regions} or
\emph{macrostates} and organizes them into a query-able structure called the
\emph{atlas}. Each macrostate is a maximal, contiguous,
nearly-equipotential-energy conformational region. In addition, EASAL uses the
novel Cayley parameters, which are a distance-based internal coordinate
representation of assembly configurations, to represent active constraint
regions, yielding convex \emph{Cayley} regions, a feature unique to assembly
energy landscapes (as opposed to folding landscapes). Convexity of the Cayley
regions improves sampling efficiency since sampling stays within the
macrostates without having to explicitly enforce constraints. Convexity also
avoids gradient descent and discarded and repeated samples which are common in
other methods.

The EASAL methodology can quickly, without relying heavily on sampling,
generate (1) a query-able atlas of local potential energy minima, their basin
structure, and energy barriers between neighboring basins; (2) paths between a
specified pair of basins, each path being a sequence of conformational regions
or macrostates below a desired energy threshold; and (3) approximations of
relative path lengths, basin volumes (configurational entropy), and path
probabilities.

The theory and algorithms behind EASAL, including results for atlas generation
decoupled from sampling, paths between basins, and approximate volume
computation with minimal sampling appear in the papers\cite{Ozkan2011, maineasal,
Ozkan2014Jacobian} and are sketched in Section \ref{sec:methods}. Preliminary
extensions of EASAL beyond dimer assemblies and viability of using EASAL for
atlasing a wide variety of assembly systems including clusters of up to 24
assembling spherical particles is demonstrated in the paper\cite{maineasal}. The
software implementation of EASAL (architecture and functionalities) is reported
in \cite{Ozkan:toms}. The software has recently been employed to predict
crucial inter-monomeric interface interactions for viral capsid assembly
\cite{Wu2012, Wu2014Virus}. 

In contrast, the primary focus of this paper is to illustrate sampling related
features of EASAL through a variety of custom-designed measurements. This is
important for EASAL's performance in tasks that call for exact sampling, such
as exact computation of basin volumes, path lengths and path probabilities.
Specifically, we compare the performance of EASAL and MC with respect to the
following sampling characteristics.

\begin{enumerate}
\item \emph{Speed:} EASAL's use of convexifying Cayley parameters allows it to
avoid gradient descent which is used in MC to enforce constraints. In addition, 
it also reduces the number discarded samples, leading to high sampling speed. 
Results comparing the sampling speed of EASAL and MC are illustrated in Section 
\ref{sec:coverage}.

\item \emph{Accuracy of uniform coverage:} To measure the accuracy of coverage, 
we need to normalize the accuracy by the number of samples. In this paper
this is measured by the $\varepsilon$-coverage (defined in Section 
\ref{sec:keyMeasurements}). Results comparing the $\varepsilon$-coverage of MC 
and EASAL are illustrated in Section \ref{sec:coverage}.

\item \emph{Efficiency of uniform coverage:} When the goal of sampling is 
the coverage of the energy landscape, an efficient sampling algorithm 
should put as few points as possible in an $\varepsilon$-region, thereby 
reducing the number of repeated samples. Results comparing the efficiency 
of uniform coverage in EASAL and MC are illustrated in Section 
\ref{sec:coverage}.

\item \emph{Accuracy of weighted coverage of lower energy regions:}
Certain applications may want lower energy regions to be sampled more densely,
compared to higher energy regions. To compare the accuracy of sampling in such 
scenarios, we use the MultiGrid (defined in Section \ref{sec:multigrid}) as 
the reference grid for comparison. Results comparing MC and EASAL in terms of
weighted coverage of lower dimensional regions are illustrated in Section 
\ref{sec:LowerDim}.

\item \emph{Localized sampling of individual active constraint regions (macrostates):}
EASAL's partitioning of the Energy landscape into active constraint regions and 
its use of convexifying Cayley parameters to represent these regions, allows it 
to localize sampling to individual active constraint regions or macrostates, of 
varying dimensions. EASAL can sample individual 5D, 4D, 3D, and 2D (with 1, 2, 3, 
and 4 active constraints respectively) regions of the energy landscape very fast, 
whereas MC has to sample the entire region to be able to locate these. Section 
\ref{sec:distribution} demonstrates EASAL's flexibility at localized
sampling of individual active constraint regions.

\item \emph{Flexibility of sampling distribution:}
EASAL's sampling flexibility permits a variety of sampling distributions, 
from uniform Cayley parameter sampling, which is skewed towards lower energy 
regions, to uniform Cartesian sampling, which is essential for accurate 
computation of configurational entropy and other integrals. This is 
illustrated in Section \ref{sec:EASALVariants}, and for all the above 
sampling attributes, we compare MC with all the different variants of EASAL.
\end{enumerate}

\eat{Accurate 
computation of configurational entropy is required for the estimation of both 
free energy and kinetics.}

\subsection{Motivation and Related Work}
\label{sec:related_work}
The problem of protein-protein assembly is an area of active research and
development \cite{ Duhovny2002, Marshall2005, Andrusier2007, Andrusier2008,
Vajda_rev2009, Dock_rev2, Dock_rev1, Dock_rev3}. Currently the most successful
approach to docking two proteins together uses a direct exhaustive search of
the whole configurational space. It is usually performed in the inverse space
for translational moves on a cubic grid (using fast Fourier transforms (FFT))
\cite{Vakser_FFT}. The FFT algorithm makes the translation search very
efficient, but it has to be repeated for all orientations of a molecule being
docked resulting in thousands of FFT operations, each comprising millions of
translations. The majority of the available software for molecular docking can
only deal with two proteins (a dimer). 

Several docking procedures are based on shape recognition and image
segmentation techniques from Computer Vision \cite{Bespamyatnikh}. The
PatchDock algorithm \cite{schneidman2005patchdock} starts with a smooth
representation of the molecular surface as a set of discrete points, but the
set is restricted to critical points (convex caps, toroidal belts and concave
pits), and the normal vectors at these points. A geometric hashing algorithm
performs a very fast matching of the caps and pits with opposing normal
directions on two surfaces, and collects all the rigid-body solutions that are
geometrically acceptable after rejecting volume overlaps. Other geometric
criteria can easily be incorporated in the procedure, for instance molecular
symmetry in SymmDock \cite{schneidman2005patchdock}, which allows building
models of oligomeric proteins with up to twenty subunits. 
Unlike the goals of these methods, which is finding site-specific docking,
the goal of EASAL and MC is sampling the entire configuration space of
the assembly landscape.

A third class of methods, including Monte Carlo (MC) simulations and molecular
dynamics (MD) are more prevalent in the study of protein folding and protein
assembly than the other two. Both these techniques sample the system's
configuration space with probabilities corresponding to the Boltzmann
distribution. Theoretically, such simulations can produce a probability density
function for the whole phase space of the system. Absolute free energy and
relative probabilities of various states in the phase space can then be
estimated. 

However, in practice, systems of interest are rarely ergodic, in the
sense that their energy landscapes consist of an unknown number of energy
minima separated by large energy barriers. Moreover, in tightly packed
molecular systems, the majority of the phase space has high energy and low
probability. In such conditions, most sampling procedures have a tendency to
over sample local basins of the energy function, and have difficulty crossing
over energy barriers. This results in uncertainty in both, i) relative
probabilities of visited states, as well as ii) whether the range of low
energy configurations visited during simulations is ever complete. Despite
recent progress all currently existing methods of protein assembly are
extremely computationally expensive.

In contrast to the more general applicability of MC and MD, the EASAL
methodology applies only to assembly. However, it leverages geometric features
unique to assembly (Cayley convexification, discussed in detail in Section
\ref{sec:backgroundEASALConvexification}) which gives it critical advantages
when sampling assembly landscapes. In addition, the EASAL methodology comes
with rigorously provable efficiency, accuracy, and tradeoff guarantees.

\noindent\textbf{Organization:} The rest of the paper is organized as follows.
Section \ref{sec:methods} briefly describes the EASAL methodology, highlighting
its salient features and different sampling distributions. Section
\ref{sec:results} compares the performance of EASAL and MC at sampling the
potential energy landscape of the two assembling transmembrane helices.
Section \ref{sec:discussion} discusses potential avenues to leverage the
complementary strengths of both the methods. Section \ref{sec:conclusion}
gives conclusions and avenues for future work.

\eat{
To overcome the problem of incomplete sampling of relevant phase space when
modeling protein assembly we use a recently introduced approach called EASAL
(Efficient Atlasing of Assembly Landscapes), for representing, visualizing,
sampling and computing integrals over the potential energy landscape tailored
for small molecular assemblies. EASAL's efficiency arises from the fact that
small assembly landscapes permit the use of so-called Cayley parameters
(inter-atomic distances) for geometric representation and sampling of constant
potential energy regions of the assembly configuration space. This results in
the isolation, convexification, customized sampling and systematic traversal
of regions using a comprehensive topological roadmap, providing reasonable
coverage of low potential energy, but narrow regions of low effective
dimension, with surprisingly few samples.

By sampling the assembly landscape of 2 Transmembrane helices, with short-range
pair-potentials, our result demonstrates that EASAL provides reasonable
coverage of crucial but narrow regions of low effective dimension with much
fewer samples and computational resources than traditional Monte Carlo or
Molecular Dynamics based sampling. Promising avenues are discussed, for
combining the complementary advantages of the two methods. 
}

\section{Materials and Methods}
\label{sec:methods}

EASAL (Efficient Atlasing and Search of Assembly Landscapes) is a novel
geometric methodology for analyzing free-energy and kinetics of assembly driven
by short-range pair-potentials in an implicit solvent.  The key concept of the
new methodology is the \emph{atlas}, which is a partition of the assembly
landscape into nearly equipotential energy regions called \emph{active
constraint regions} or \emph{macrostates}. The atlas is organized as a
refinable, and query-able roadmap, which gives neighborhood relationships
between active constraint regions.  The active constraint regions have unique
labels, which are graphs of active geometric constraints called \emph{active
constraint graphs}. The constraints are the pairwise Lennard-Jones potentials
which drive assembly, along with sterics. Atom pairs (one from each
transmembrane helix) are said to be an \emph{active constraint} when the
distance between their centers are in the Lennard-Jones's well.  The active
constraint graphs are analyzed using geometric rigidity techniques, whereby the
energy level of the active constraint region becomes a proxy for its dimension.

The EASAL methodology produces: (1) a query-able atlas of local potential
energy minima, their basin structure, and energy barriers; and (2)
approximations of basin volumes (configurational entropy).  Leveraging its
rigorous geometric underpinnings, EASAL  can generate the topological roadmap
of the energy landscape without relying heavily on sampling \cite{maineasal}.
In addition, EASAL provides provable guarantees for tradeoffs between
efficiency and accuracy.
	
The input to EASAL is an assembly system consisting of the following
\begin{enumerate}
\item A collection of $k$ \emph{rigid molecular components},  
each with at most $n$ atoms in them. The rigid molecular components are specified as
the positions of \emph{atom-centers} and \emph{atom radii}, in a local coordinate system.

\item Pairwise component of the potential energy function, specified using \emph{Lennard-Jones} 
\emph{potential energy terms} (subsuming \emph{Hard-Sphere} potential).
The pairwise Lennard-Jones term for a pair of atoms, $i$ and $j$, one from each 
rigid molecular component, 
is given as a function of the distance $d_{i,j}$ between the centers of $i$ and $j$. 
The \emph{atom-center} could, in some cases, be the representation
for the average position of a \emph{collection of atoms in a residue}.

\item A non-pairwise component of the potential energy function in the form of 
\emph{global potential energy} terms that capture other factors including the 
implicit solvent (water or lipid bilayer membrane) effect \cite{Lazaridis_Karplus_1999, 
Lazaridis_2003, Im_Feig_Brooks_2003}. These are specified as a function of the entire 
assembly configuration.

\item An optional set of constraints of interest, specified as a set of atom
pairs, one from each rigid molecular component.  When specified, only those 
configurations, which contain active constraints from the specified pairs of 
atoms are sampled. 

\item The desired level of refinement of sampling is specified as the 
\emph{sampling step size} $t$.
\end{enumerate}

To sample the configuration space of an assembly system with input as described
above, EASAL employs several key strategies including geometrization,
stratification, recursive exploration of lower dimensional regions, and Cayley
convexification.  Each of these is described next.

\subsection{Geometrization}
\label{sec:backgroundEASALSGeometrization}
The input Lennard-Jones pair potentials are \emph{geometrized}, into distance
constraints.  For the inter-atomic short-range Lennard-Jones potential energy,
the distance between the centers of atoms, are discretized into 3 main regions,
large distances at which Lennard-Jones potentials are no longer relevant, short
distances where inter-atomic repulsion kicks in and the interval between the
two known as the \emph{Lennard-Jones' well}.

\subsection{Stratification}
\label{sec:backgroundEASALStratification}
EASAL partitions the energy landscape of the assembling transmembrane helices
into active constraint regions.
The active constraint regions are organized as a roadmap that captures their
boundary relationships. In particular, the active constraint graph of a region
is a subgraph of the active constraint graph of its children boundary regions. 
The children of a given region are 1 lower dimensional regions obtained when
one additional constraint becomes active.

\begin{figure}[htpb]
   \centering
	\includegraphics[width=0.75\linewidth]{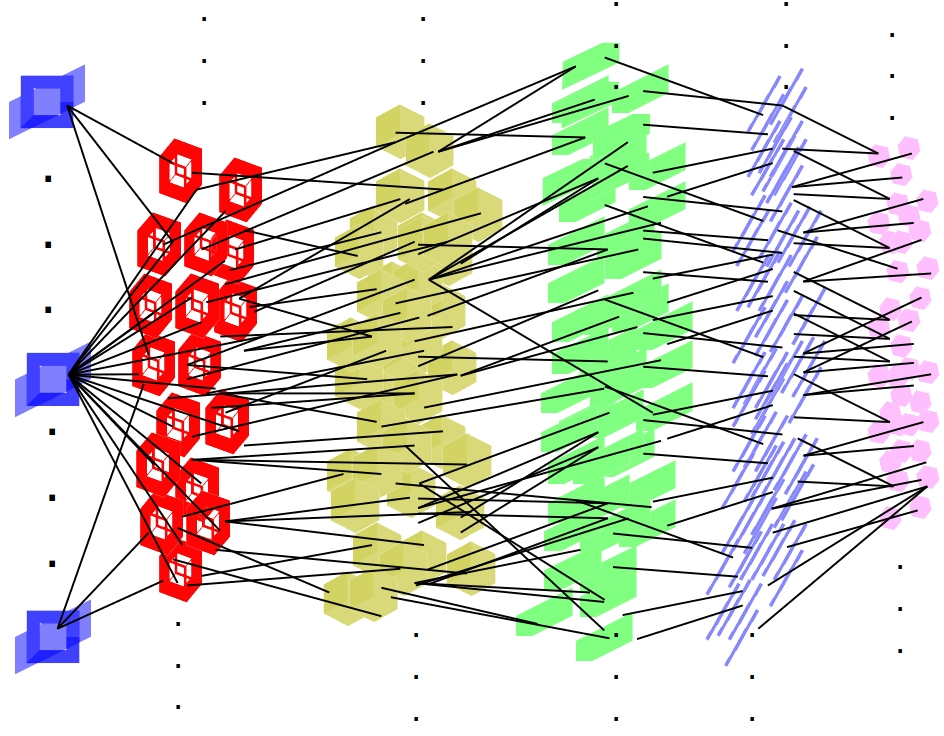}
	\includegraphics[scale=0.2]{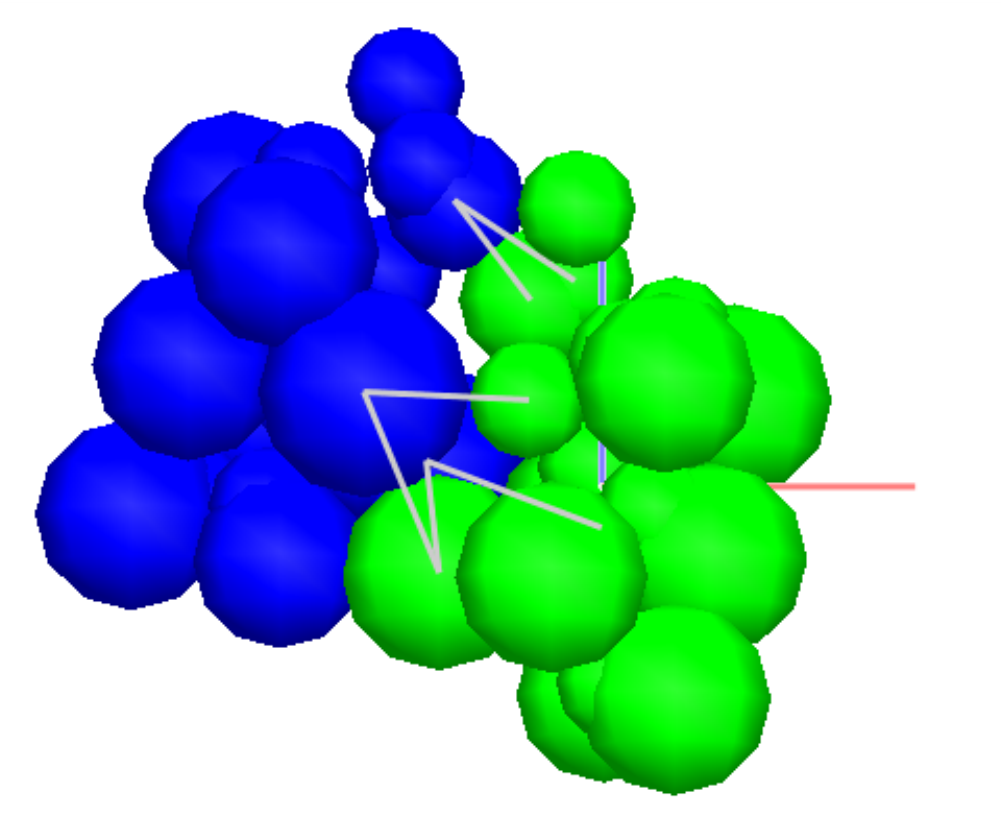}
  \caption{A portion of the stratified roadmap for the input
		molecules shown at the bottom right. The Active
		constraints regions of various dimensions are shown in different
		colors, 5D (blue), 4D (red), 3D (yellow), 2D (green), 1D (purple), 0D (pink).
		Edges indicate containment in a parent region one dimension higher.
		See Section \ref{sec:backgroundEASALStratification}.}
      \label{fig:atlas}
\end{figure}

Configurations with more active constraints have lower potential energy and the lowest
potential energy is attained at the bottoms of the potential energy basins.
This leads to a dimensional stratification of the active constraint regions
called a Thom-Whitney Stratification \cite{springerlink:10.1007/BF01420960}.
Intuitively, the active constraint regions are put in strata, one for each
dimension and each stratum consists of active constraint regions of the same
effective dimension. Figure \ref{fig:atlas} shows a screenshot from the EASAL
software, depicting an atlas.

\subsection{Recursive exploration of lower dimensional active constraint regions}
\label{sec:backgroundEASALRecursiveExploration}
EASAL uses a recursive method that starts exploration from the interiors of
higher dimensional regions and searches for the presence of boundary regions of
exactly one lesser dimension (with exactly one extra active constraint).
Searching for boundaries one dimension less at every stage, has a higher chance
of success than looking for the lowest dimensional active constraint regions
directly.

When a new child region is discovered, all its higher dimensional ancestor
regions are immediately discovered since they correspond to active constraint
regions with a subset of the active constraints.  So, even if a small
(hard-to-find) region is missed at some stage (due to the sampling step size
being too large), if any of its descendants are found at a later stage, say via
a larger (easy-to-find) sibling, the originally missed region is discovered.
This feature is especially useful when dealing with large molecules with
intricate topologies, since finding these missed regions would otherwise
involve sampling with a smaller step size, thereby increasing the atlasing
time.

\subsection{Convexification}
\label{sec:backgroundEASALConvexification}
\begin{figure}[htpb]
		\includegraphics[height=3in, width=0.7\linewidth]{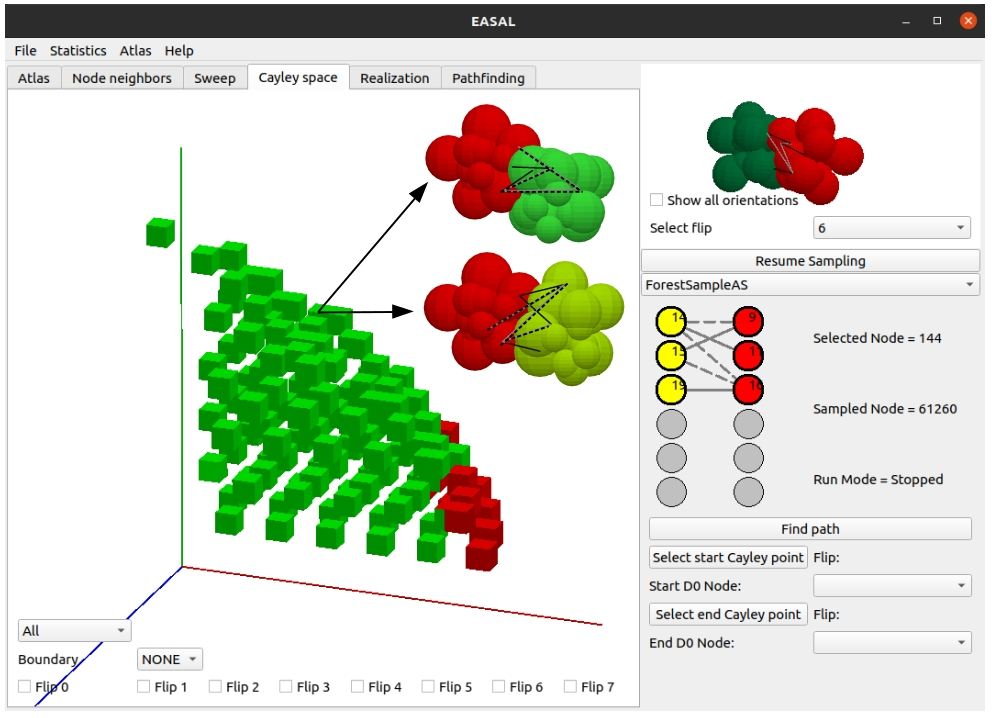}
	\caption{A screenshot of the EASAL software. The main screen shows the Cayley configuration
		space of an active constraint region, whose active constraint graph
		is shown on the bottom right. Green points represent valid configurations
		and the red points represent configurations that have streic collisions.
		We show two Cartesian configurations of the same \Cp.
		The solid edges on the shown Cartesian configurations represent active constraints
		between the atoms they connect and the dotted edges represent the chosen Cayley parameters. 
		See Section \ref{sec:backgroundEASALConvexification}.}
   \label{fig:prtree}
\end{figure}

To find lower dimensional boundaries, EASAL uses the theory of convex
\emph{Cayley} (distance-based) parameterization \cite{SiGa:2010}. EASAL
efficiently maps (many to one) a $d$-dimensional active constraint region to a
convex region of $\mathbb{R}^d$ called the Cayley space of the region (see
Figure \ref{fig:prtree}). Sampling in a convex space is more efficient as it
alleviates the need for gradient-descent search to enforce constraints, which is
required in MC and other prevalent methods.  This significantly reduces the
number of repeated and discarded samples and is one of the main reasons for
EASAL's atlasing efficiency. In addition, it is extremely easy to
compute the inverse map from the samples in the Cayley space to their
corresponding finitely many configurations in the Cartesian space.  This
parameterization leverages geometric features that are  unique to assembly (as
opposed to protein folding).
\subsection{EASAL Variants and their Sampling Distributions} 
\label{sec:EASALVariants}
EASAL's flexibility is demonstrated with a variety of sampling distributions
in the Cayley space, which translates to sampling variants in the Cartesian
space. EASAL-1 samples the Cayley space uniformly, skewing the distribution
towards lower dimensional or lower energy regions. EASAL-2 uses a step size
inversely proportional to the Cayley parameter value, resulting in more dense
sampling in the interiors of active constraint regions.
EASAL-3 uses a step size linearly proportional to the Cayley parameter value,
resulting in dense sampling close to the boundaries of active constraint
regions. EASAL-Jacobian uses a sophisticated Cayley sampling method
\cite{Ozkan2014Jacobian} to force uniform sampling in the Cartesian space. It
recursively and adaptively computes the next Cayley step size and direction
using a computation of the Jacobian of the Cartesian-Cayley map
to achieve this goal. Figure \ref{fig:easal-variants} shows the effect of
sampling using different variants of EASAL on a typical randomly chosen 2D active
constraint region.

\def\widss{0.2\linewidth}
\def\widsss{0.18\linewidth}
\begin{figure}[htpb]
   \centering
   \begin{subfigure}[b]{\widss}
      \epsfig{file=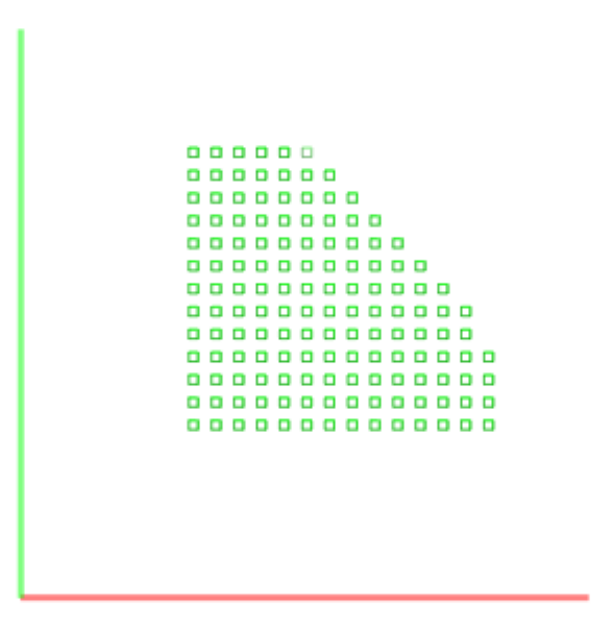, width=\linewidth}
      \caption{EASAL-1}
      \label{fig:easal1_cayley}
   \end{subfigure}
  \begin{subfigure}[b]{\widss}
      \epsfig{file=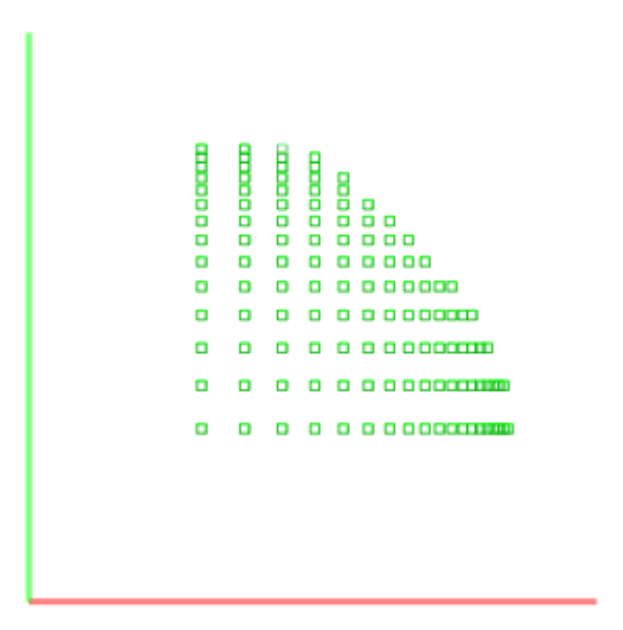, width=\linewidth}
      \caption{EASAL-2}
      \label{fig:easal2_cayley}
   \end{subfigure}
  \begin{subfigure}[b]{\widss}
      \epsfig{file=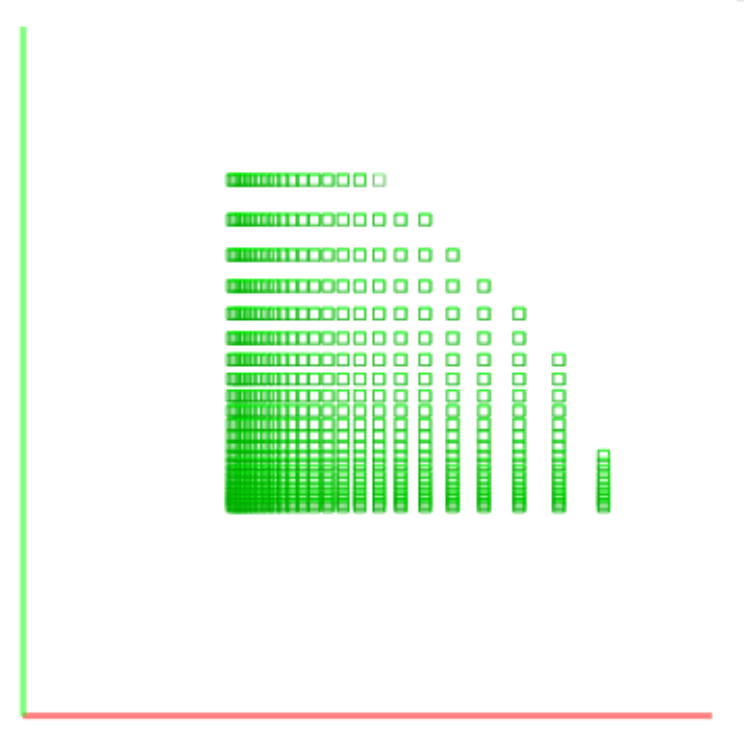, width=\linewidth}
      \caption{EASAL-3}
      \label{fig:easal3_cayley}
   \end{subfigure}
  \begin{subfigure}[b]{\widss}
      \epsfig{file=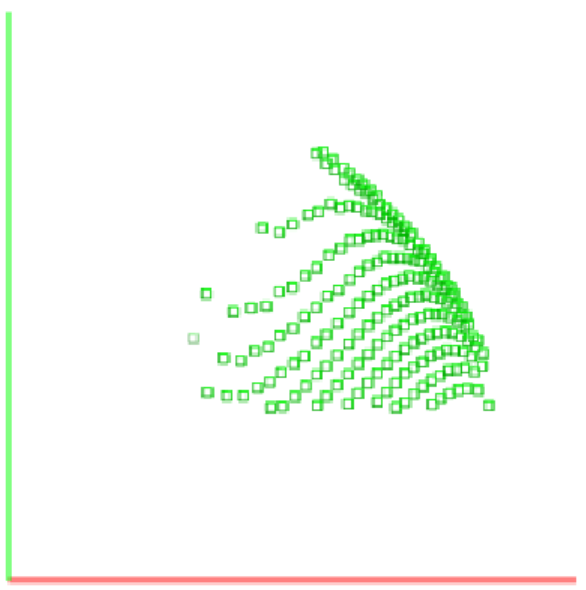, width=\linewidth}
      \caption{EASAL-Jacobian}
      \label{fig:easalJacobian_cayley}
   \end{subfigure}
  \begin{subfigure}[b]{\widss}
      \epsfig{file=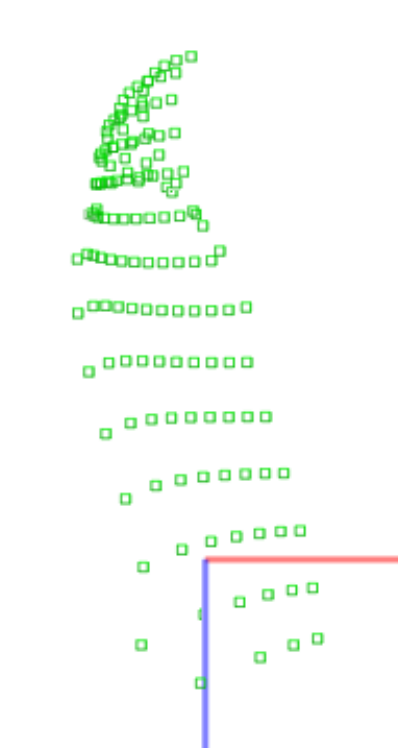, width=\linewidth}
      \caption{EASAL-1}
      \label{fig:easal1_cartesian}
   \end{subfigure}
  \begin{subfigure}[b]{\widss}
      \epsfig{file=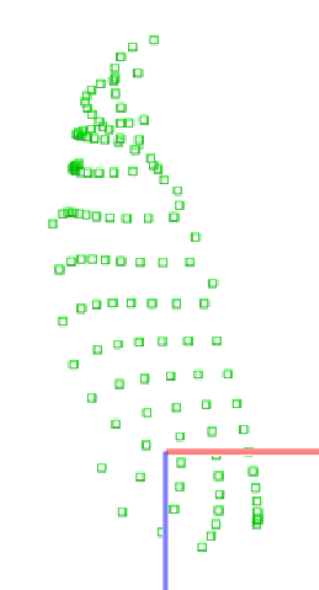, width=\linewidth}
      \caption{EASAL-2}
      \label{fig:easal2_cartesian}
   \end{subfigure}
  \begin{subfigure}[b]{\widss}
      \epsfig{file=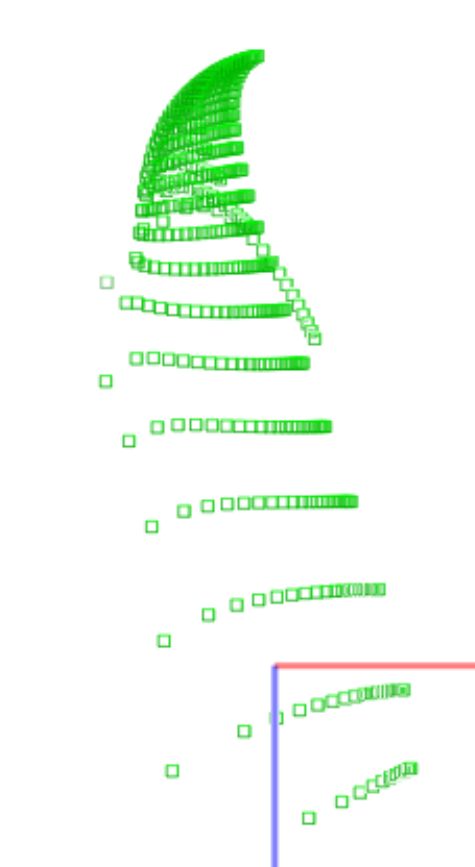, width=\linewidth}
      \caption{EASAL-3}
      \label{fig:easal3_cartesian}
   \end{subfigure}
  \begin{subfigure}[b]{\widss}
      \epsfig{file=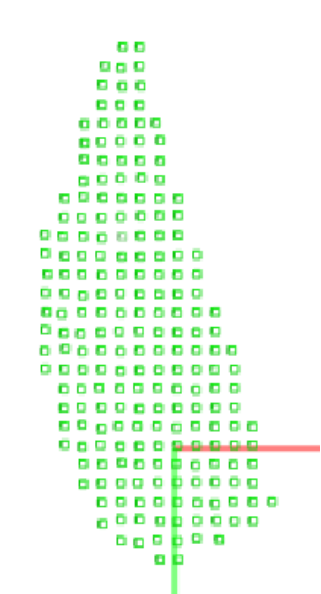, width=\linewidth}
      \caption{EASAL-Jacobian}
      \label{fig:easalJacobian_cartesian}
   \end{subfigure}
		\caption{ Comparison of sampling distribution in Cayley (top row) v/s. Cartesian space
		(bottom row) in variants of EASAL for a 2D active constraint region in the atlas with 
		the example in Figure \ref{fig:atlas} as input.  The axes in the top figures are
		the two Cayley parameters. In the bottom figures, the projection is on
		the $xy$ coordinates of the centroid of the second point-set with the
		centroid of the first point-set fixed at the origin.
		   See Section \ref{sec:EASALVariants}.}
 \label{fig:easal-variants}
\end{figure}

\section{Results}
\label{sec:results}
Section \ref{sec:expSetup} describes the experimental setup. 
Section \ref{sec:keyMeasurements} discusses the key measurements
used to compare the different sampling algorithms in this paper. Section
\ref{sec:coverage} discusses the performance of the sampling algorithms at
uniform coverage, in terms of their speed, efficiency and accuracy.
Section \ref{sec:LowerDim} discusses their performance at
accurately covering lower dimensional regions. Section \ref{sec:distribution} 
illustrates the advantages of EASAL's flexibility in terms of being able
to restrict sampling to specified macrostates.

\subsection{Experimental Setup}
\label{sec:expSetup}
We compare the performance of five sampling algorithms, MC,
EASAL-1, EASAL-2, EASAL-3, and EASAL-Jacobian with 
two identical transmembrane helices, each with 11
residues and 20 atoms as input (see Figure \ref{fig:atlas}).

\subsubsection{Metropolis Monte Carlo Experimental Setup} 
\label{sec:MC}
Metropolis Monte Carlo (MC) is an importance sampling method that generates an
ensemble according to the Boltzmann factor. MC simulations were performed in
order to explore the conformational space that is accessible by translational
and rotational random steps of rigid helices. In all simulations reported,
protein transmembrane alpha-helices were held rigid. The rigid body MC
simulation was implemented in the HARLEM program \cite{kurnikov1999harlem} 
on a machine with Intel(R) Core(TM) i5-2540 CPU.

\noindent\textbf{Move Sets:}
\label{sec:MoveSets}
Trial conformations of rigid bodies are generated by a basic move set of a
small translational and rotational displacement. The maximum step size for both
types of displacements follow a well known criterion of the MC acceptance
ratio. This criterion establishes that for optimum sampling the acceptance
ratio should be around 50\% of the trial moves. However, from our
experience this high ratio of acceptance generates random structures with small
conformational fluctuations. In order to overcome high energy barriers we
implemented a move set based on the exponential distribution of the maximum
step size of translation and rotation. The purpose of this move set is to
randomly generate trial conformations with higher probability to jump over the
energy barrier.

Moreover, analyzing MC trajectories showed that low energy conformations of
different energy basins are conformationally different by the rotation around
the principal axis of a rigid helix. Using this information we included in our
MC implementation a move set in which a helix is randomly rotated around its
principal axis.

\noindent\textbf{Scoring Energy:}
\label{sec:ScoringEnergy}
The inter-molecular energy $E_{int}$ of a structure model in this work is 
calculated as follows:
\begin{equation}
\label{eq:scoringEnergy}
E_{int} = w_1 E_{pairwise} + w_2 E_{steric} + w_3 E_{mem} + w_4 E_{vol} + w_5 E_{solvation} 
\end{equation}
where $E_{pairwise}$ is the pairwise distance-based potential of the mean force
of interaction between two residues, $E_{steric}$ is the steric overlap energy,
$E_{mem}$ is the energy term that constrains the transmembrane helices to the
membrane plane, $E_{vol}$ prevents the helix mass center from sampling farther
than a radius of 15 \AA, and $E_{solvation}$ accounts for the
interaction of amino acids in different regions of the lipid bilayer. 

\subsubsection{EASAL Experimental Setup}
With $d_{ij}$, the distance between the centers of atoms $i$ and $j$, and $r_i$
and $r_j$, the radii of atoms $i$ and $j$ respectively, we set the input
pairwise distance component, for EASAL, as follows. For all atom pairs $(i,j)$
belonging to different rigid molecular components, the steric collision distance was
set to $d_{ij} > 0.8 * (r_i + r_j)$, the distance beyond which Lennard-Jones
interactions are no longer relevant was set to $d_{ij} < r_1 + r_2 + 0.9$, and
the distance between these two is the Lennard-Jones well.  The EASAL results
were generated using its the curated opensource, software implementation,
\cite{Ozkan:toms} (software available at
\url{http://bitbucket.org/geoplexity/easal}, see also video
\url{https://cise.ufl.edu/~sitharam/EASALvideo.mpeg}, and user guide
\url{https://bitbucket.org/geoplexity/easal/src/master/CompleteUserGuide.pdf}).
The EASAL experiments were run on an Intel(R) Core(TM) 2 Quad Q9450 @ 2.66GHz
CPU with 3.9 GB of RAM. With this setup, sampling with EASAL1 took 3 hours 8
minutes, EASAL2 took 4 hours 24 minutes, EASAL3 took 10 hours 20 minutes, and
EASAL-Jacobian took 14 hours 22 minutes.

\subsubsection{Reference Grid for Uniform Coverage}
\label{sec:grid}
In our experiments, uniform coverage is measured over a uniform grid in the
Cartesian space. The angle parameters in the Cartesian space are represented in
Euler angles (ZXZ). The grid is bounded along the
6 different Cartesian directions as follows, $X, Y : [-20\si{\angstrom},
20\si{\angstrom}]$, $Z : [-3.5\si{\angstrom}, 3.5\si{\angstrom}]$ $\phi, \psi :
[-\pi, \pi]$, and the inter principal-axis angle $\theta :[0^\circ, 30^\circ]$,
where $\theta = a\cos(uv)$ with $u$ and $v$ the principal axis of the two
helices. 

\subsubsection{MultiGrid: A Reference Grid for Weighted Coverage}
\label{sec:multigrid}
To compare the performance of the sampling algorithms at being able to densely
sample lower energy regions, we define a new reference grid, called
\emph{MultiGrid}. In MultiGrid, each grid point with $l$ active constraints
(pairs of atoms in the Lennard-Jones' well) is weighted by $l$, thus favoring
lower dimensional, lower energy regions.

\subsection{Key Measurements}
\label{sec:keyMeasurements}
We describe the key measurements used to compare MC with EASAL variants 
for the different sampling characteristics described in the introduction.
\subsubsection{Measurements for Uniform Coverage}
\label{sec:keyMeasurements:UniformCoverage}
We measure the accuracy, efficiency, and speed of the sampling algorithms
for covering the uniform grid described in Section \ref{sec:grid}. To measure
the speed, we use the \emph{number of samples} required by the sampling
algorithms.

\begin{figure}[htpb]
\includegraphics[scale=0.5]{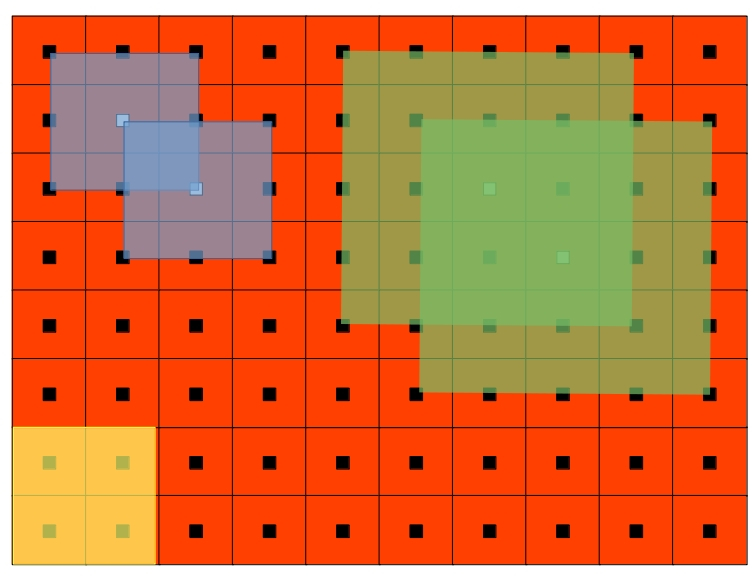}
\caption{Illustration of epsilon cubes on a 2 dimensional grid. Black dots
		represent grid points, each of which is one step size away from its
		neighbor in every dimension.  Red squares around the grid points are
		grid cubes.  The shaded blue and green areas represent
		$\varepsilon$-cube and $\varepsilon+1$-cubes respectively, centered
		around the grid point in the middle of the cube (highlighted in white
		instead of black). The $\varepsilon$ and the $\varepsilon+1$ cubes
		extend 1 and 2 step sizes respectively from the center grid point in
		every dimension. As can be seen, there is significant overlap between
		epsilon cubes of neighboring points. The shaded yellow area represents
		a coarse grid cube. See Section \ref{sec:keyMeasurements:UniformCoverage}} 
\label{fig:epsilonRegions} 
\end{figure}

The accuracy of the different sampling algorithms at uniform coverage is
measured using the \emph{epsilon coverage}. An $\varepsilon$-cube is a cube
centered around a grid point with a range of $2\varepsilon$ in each of the 6
dimensions (see Figure \ref{fig:epsilonRegions}). The accuracy of uniform
coverage is the percentage of $\varepsilon$-cubes covered by at least one
sample point.

Since each of the sampling algorithms uses different number of samples, 
accuracy measures need to be normalized by the number of samples. 
We do this by defining the $\varepsilon$-cube for a sampling algorithm
based on the number of samples. The value
of $\varepsilon$ is set to: $$\varepsilon := \lceil
\frac{\sqrt[6]{\frac{\Gamma}{\sigma}}}{2}\rceil$$ where $\Gamma$ is the total
number of grid points and $\sigma$ is the number of EASAL/MC sample points. To
compute $\varepsilon$-coverage, we assign each EASAL/MC sample to its closest
grid point. Call those grid points \emph{EASAL/MC-mapped} grid points. We say
that a grid point $p$ is \emph{covered} by EASAL/MC if there is at least one
EASAL/MC -mapped grid point within the $\varepsilon$-cube centered around $p$.
The better sampling algorithm will have greater $\varepsilon$-coverage.

Efficient coverage requires sampling algorithms to minimize repeated samples by
using as few samples as possible to cover an $\varepsilon$-cube (ideally only
1). We define two measures of efficiency. The first $\mu_1$, is the ratio of
$\varepsilon$-cubes sampled efficiently (with exactly 1 sample) to the ratio of
$\varepsilon$-cubes sampled inefficiently (with more than 1 samples). The
second $\mu_2$, is the ratio of the number of $\varepsilon$-cubes sampled
efficiently (with exactly 1 sample) to the total
number of samples in all other cubes.

We measure the speed of the different sampling algorithms using the
\emph{number of samples} required to achieve a certain $\epsilon$-coverage. The
better algorithm uses fewer number of samples. 

\subsubsection{Measurements for Weighted Coverage of Lower Energy Regions}
\label{sec:keyMeasurements:WeightedCoverage}
To measure the accuracy of weighted coverage with respect to lower energy regions,
we compare the distribution of samples in the different sampling algorithms to 
the distribution of samples in the reference MultiGrid.

A key measure here is the \emph{distribution ratio} of an
active constraint region, which is the ratio of the number of samples in the
region to the total number of samples in all regions in the atlas. We compare
the distribution ratios of EASAL variants and MC to MultiGrid.

\subsubsection{Measurements for Localized Sampling of Individual Constraint Regions}
To demonstrate EASAL's sampling flexibility which allows individual active
constraint regions to be sampled quickly, we compare the number of samples
required to sample a macrostate to the total number of samples in all
macrostates.  Since MC does not have the ability to restrict sampling to
macrostates, in the worst case, it may need to sample the entire energy
landscape to sample a single macrostate. 

Let $s_1$ be the number of samples in a specific but randomly chosen
three-dimensional region and $s_2$ be the number of samples in all ancestor
regions with one active constraint that lead to the three-dimensional region.
The \emph{ratio percentage} is $\frac{s_1}{s_2} \times 100$. Ratio percentage
is a measure of the percentage of samples in an ancestor 5-dimensional node
that will end up in a given 3-dimensional active constraint region. 

\subsection{Uniform Coverage: Accuracy, Efficiency, and Speed}
\label{sec:coverage}
In this experiment, we compare the performance of the different sampling
algorithms at covering the uniform grid described in Section \ref{sec:grid}.
The best method should have the
highest epsilon coverage with the fewest number of samples.
Table \ref{table:coverage} summarizes the coverage results, in terms 
of speed and accuracy for MC and 4 EASAL
variants. MC gives the best coverage at $99.96\%$, but requires 100 million
samples to do the same. Compare this to EASAL-Jacobian, which gives similar
coverage with $99.53\%$ with a hundredth the number of samples as MC. EASAL-2
gives a very good coverage of 92.42\% with only 40k samples (0.04\% of MC).

\begin{table*}[htpb]
\centering
\caption{Comparison of MC with EASAL variants for speed and accuracy of uniform coverage, 
with the two transmembrane helices shown in \figref{fig:atlas} as input.
See Section \ref{sec:coverage}.}
\label{table:coverage}
\resizebox{\textwidth}{!}{%
\begin{tabular}{lccccccc}\hline
		sampling method & EASAL-1 &EASAL-2& EASAL-3& EASAL-Jacobian& MC &Grid &MultiGrid\\\hline
		$\varepsilon$ value& $\lceil 0.97\rceil$ & $\lceil 1.14\rceil$& $\lceil 1.20\rceil$& $\lceil 0.66\rceil$ &$\lceil0.31\rceil$&N/A&N/A\\\hline
		$\varepsilon$ coverage &92.06\% &92.42\% &74.08\% &99.53\% &99.96\%&N/A&N/A\\\hline
		Number of Samples & 100k & 40k & 30k & 1 million & 100 million&6 million&12 million\\\hline
\end{tabular}
}
\end{table*}

\begin{figure*} [htbp]
\def\wid{.35\textwidth}
\centering
 \begin{subfigure}[b]{\wid}
      \epsfig{file = 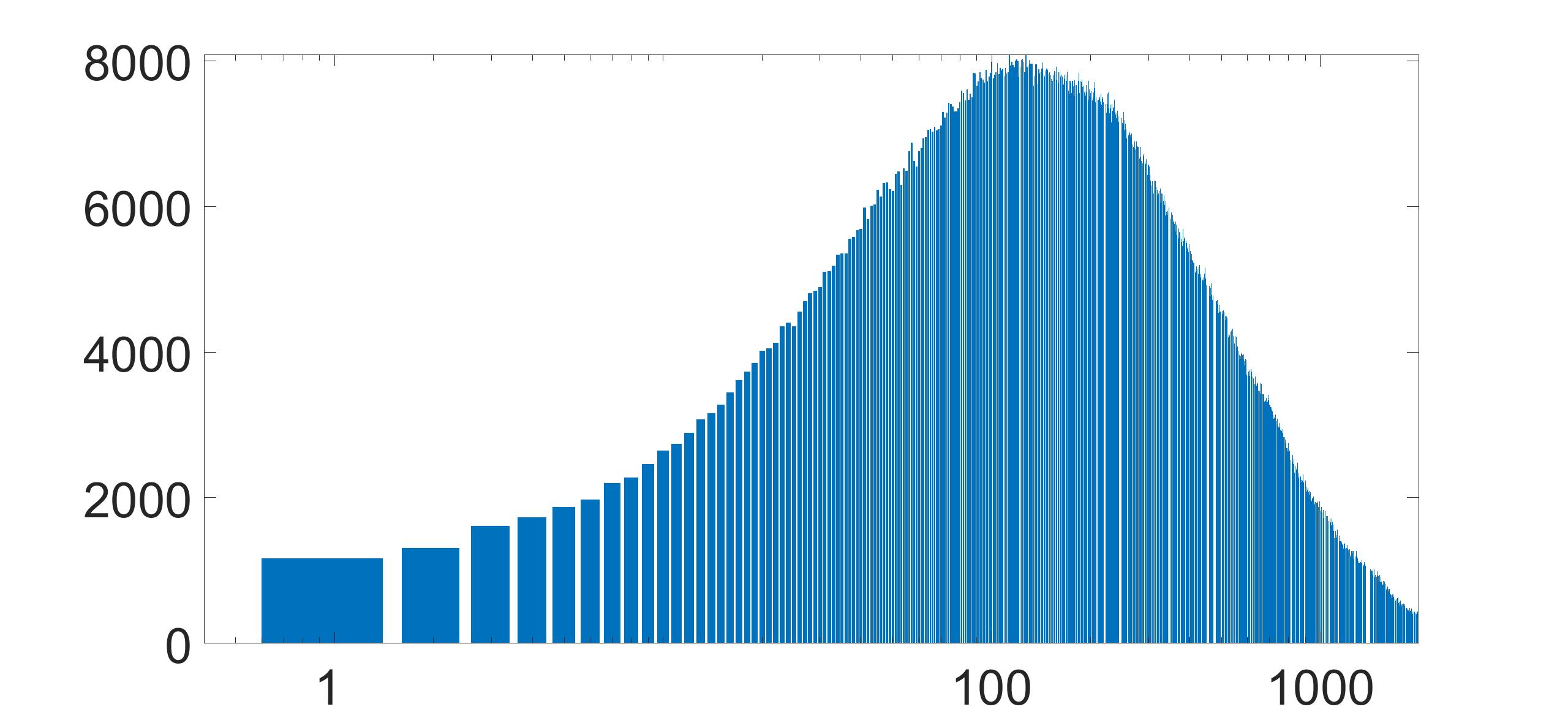, width=\linewidth}
      \caption{MC}
   \end{subfigure}
 \begin{subfigure}[b]{\wid}
      \epsfig{file = 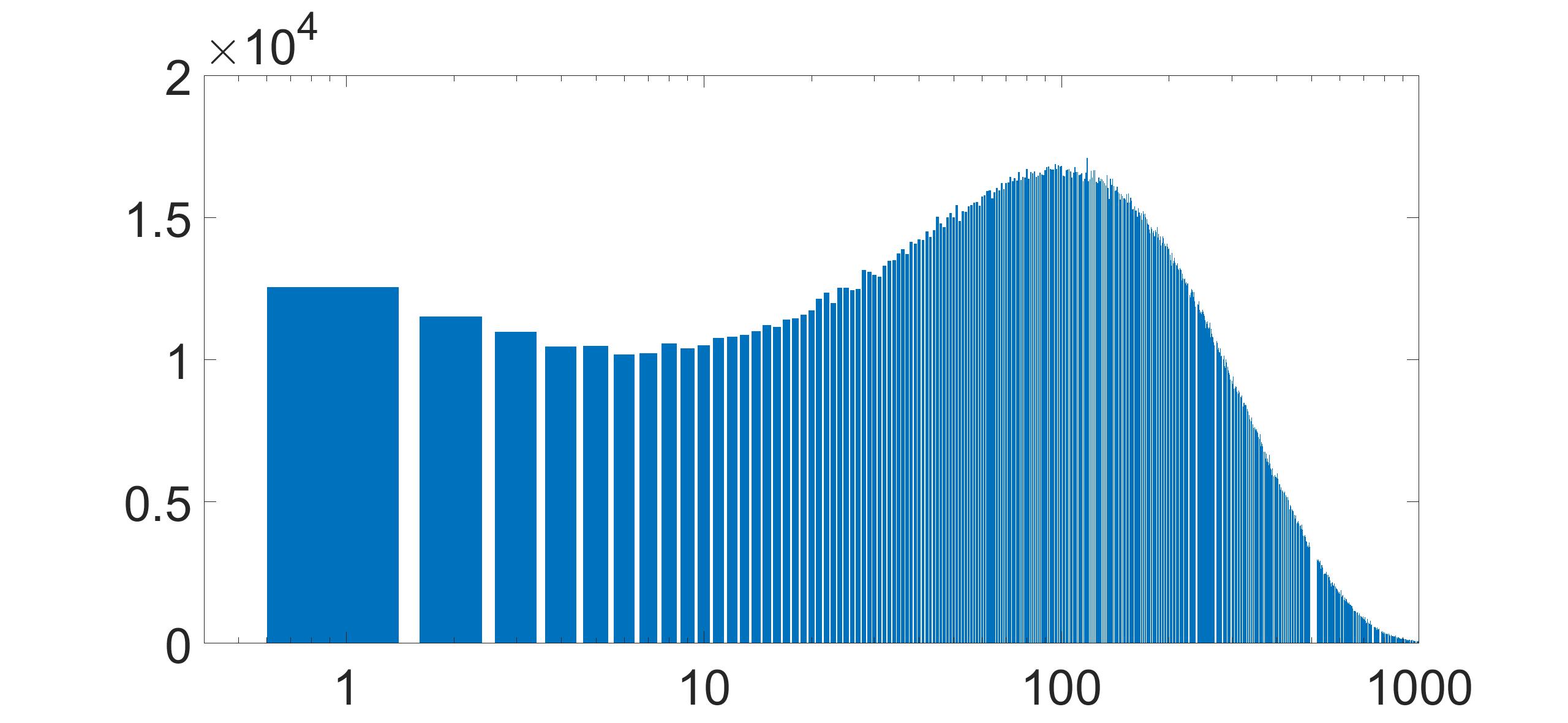, width=\linewidth}
      \caption{EASAL-Jacobian}
   \end{subfigure}
 \begin{subfigure}[b]{\wid}
      \epsfig{file = 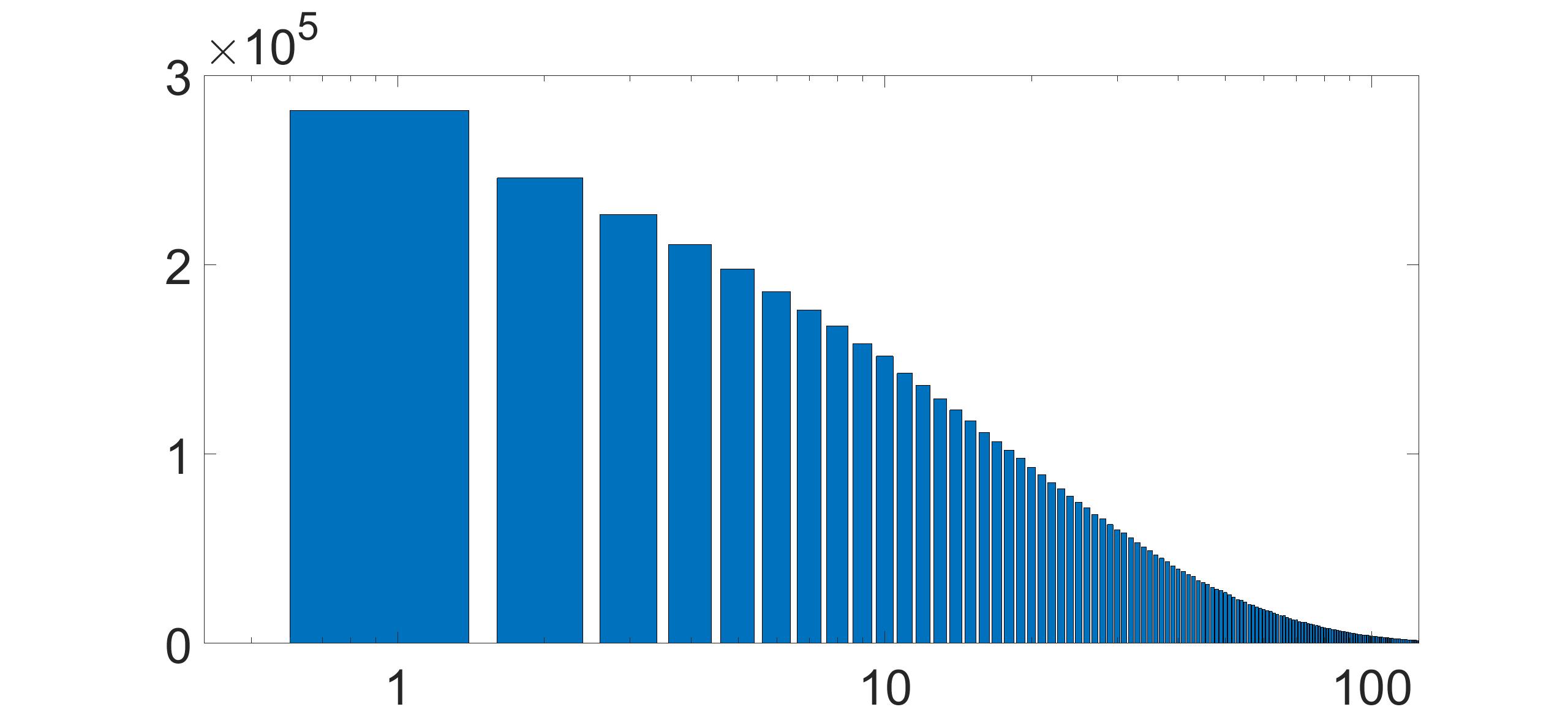, width=\linewidth}
      \caption{EASAL-1}
   \end{subfigure}
 \begin{subfigure}[b]{\wid}
      \epsfig{file = 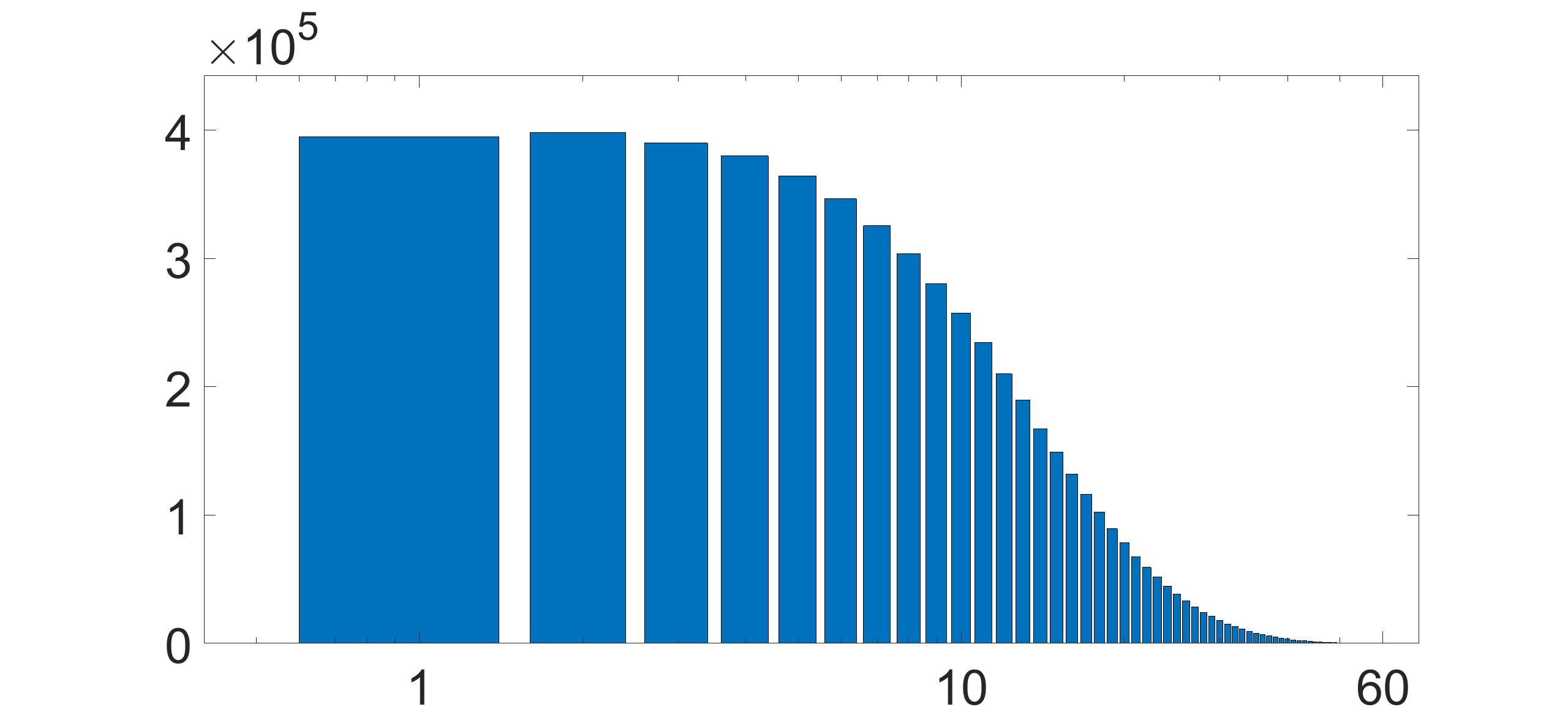, width=\linewidth}
      \caption{EASAL-2}
   \end{subfigure}
 \begin{subfigure}[b]{\wid}
      \epsfig{file = 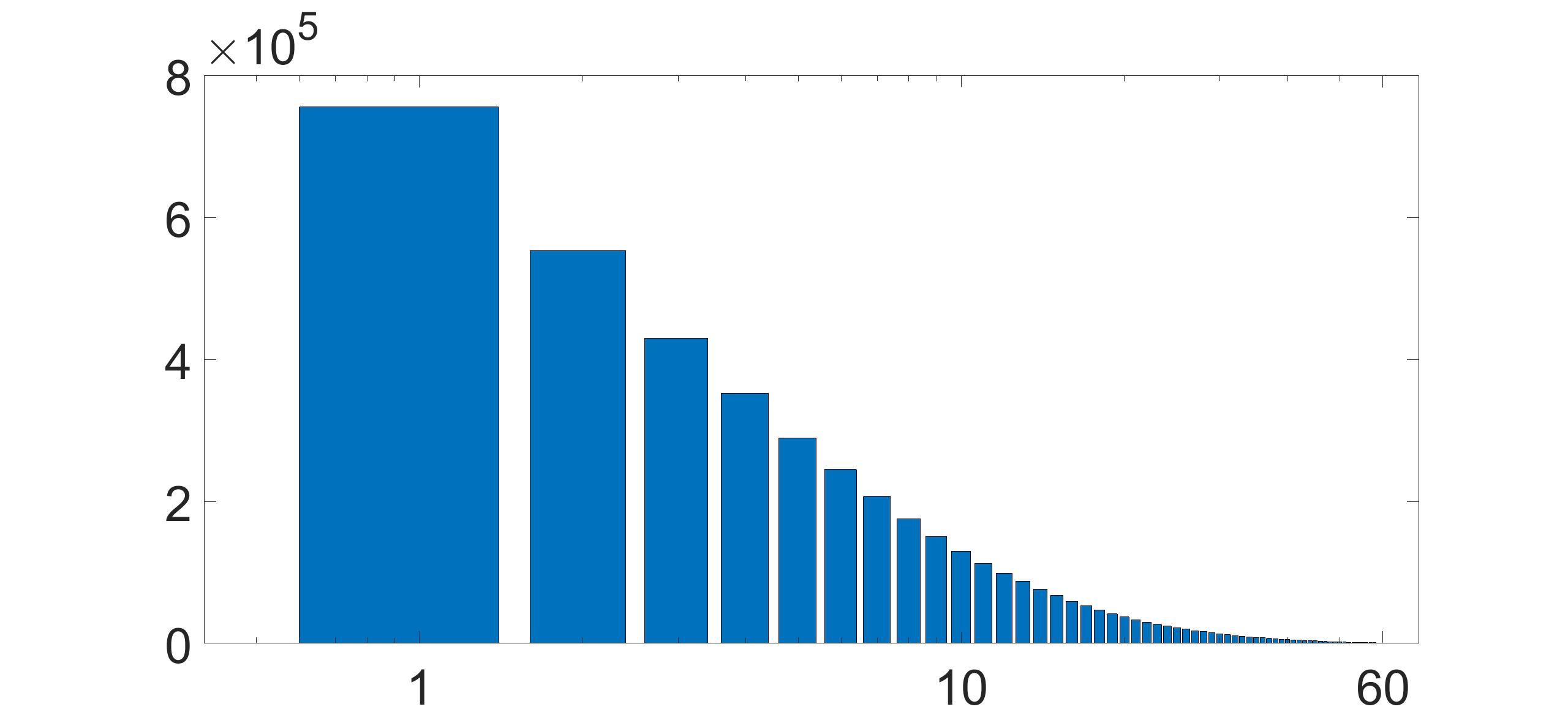, width=\linewidth}
      \caption{EASAL-3}
   \end{subfigure}
\caption{The horizontal axis shows the number of sample points $\nu$ that lie in $\varepsilon$-cubes,
and the vertical axis shows the number of $varepsilon$-cubes having $\nu$ EASAL/MC-mapped points inside of them.
See Section \ref{sec:coverage}.
}
\label{fig:PointsVsRegions}
\end{figure*}

To compare the sampling algorithms in terms of efficiency at covering the 
uniform grid, Figure \ref{fig:PointsVsRegions} plots the number of sample points $\nu$ (from
the different sampling algorithms) that lie in an $\varepsilon$-cube against the
number of $\varepsilon$-cubes having $\nu$ EASAL/MC-mapped, sample points in them. A good
sampling algorithm (one that minimizes repeated samples) should have the fewest
number of sample points per $\varepsilon$-cube mapped to the same epsilon cube. In the
plot, this would appear as a histogram skewed heavily to the left. As can be
seen, EASAL-1, EASAL-2, and EASAL-3 have peaks at the left-most part of the
histogram, indicating that they minimize discarded samples better than MC and
EASAL-Jacobian.

\begin{figure*} [htbp]
\def\wid{.35\textwidth}
\centering
 \begin{subfigure}[b]{\wid}
      \epsfig{file = 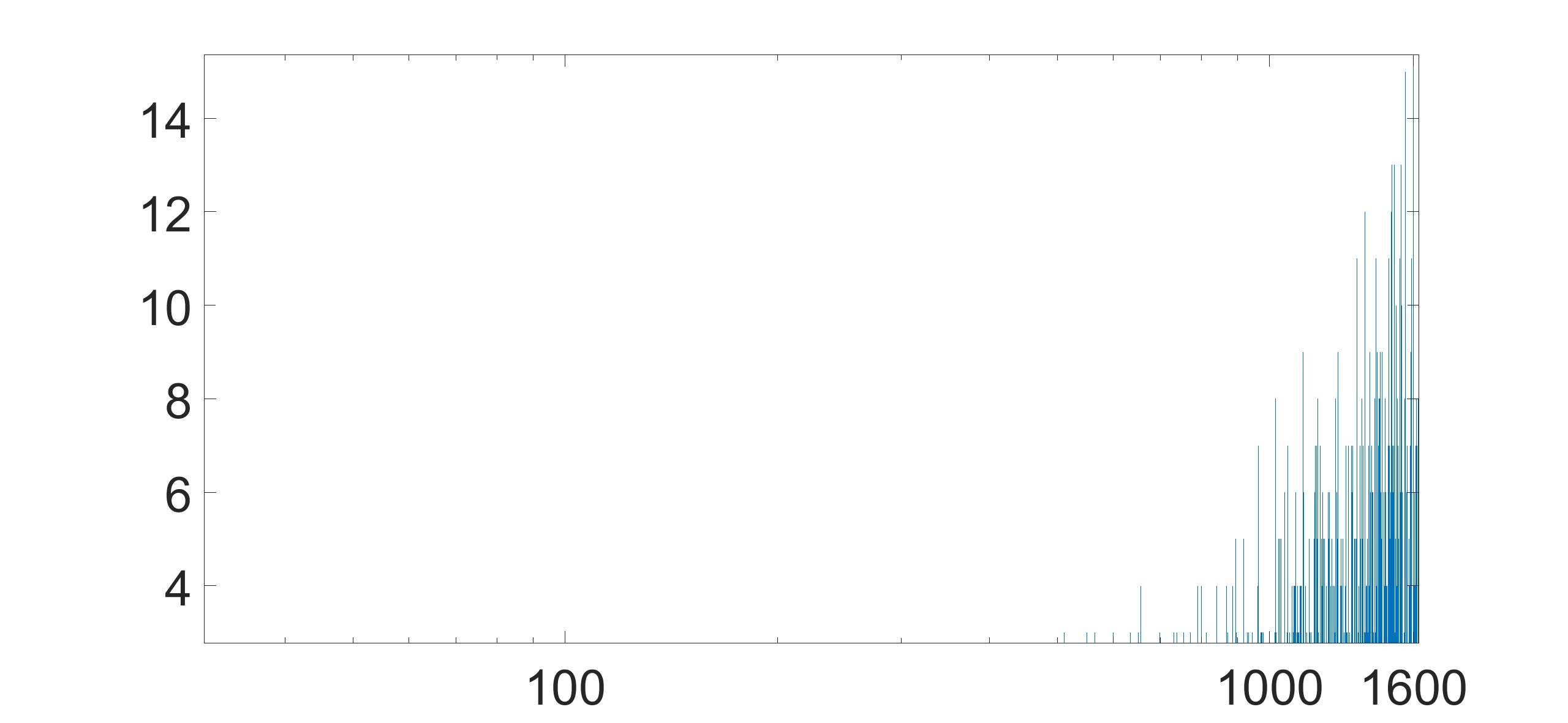, width=\linewidth}
      \caption{MC}
   \end{subfigure}
 \begin{subfigure}[b]{\wid}
      \epsfig{file = 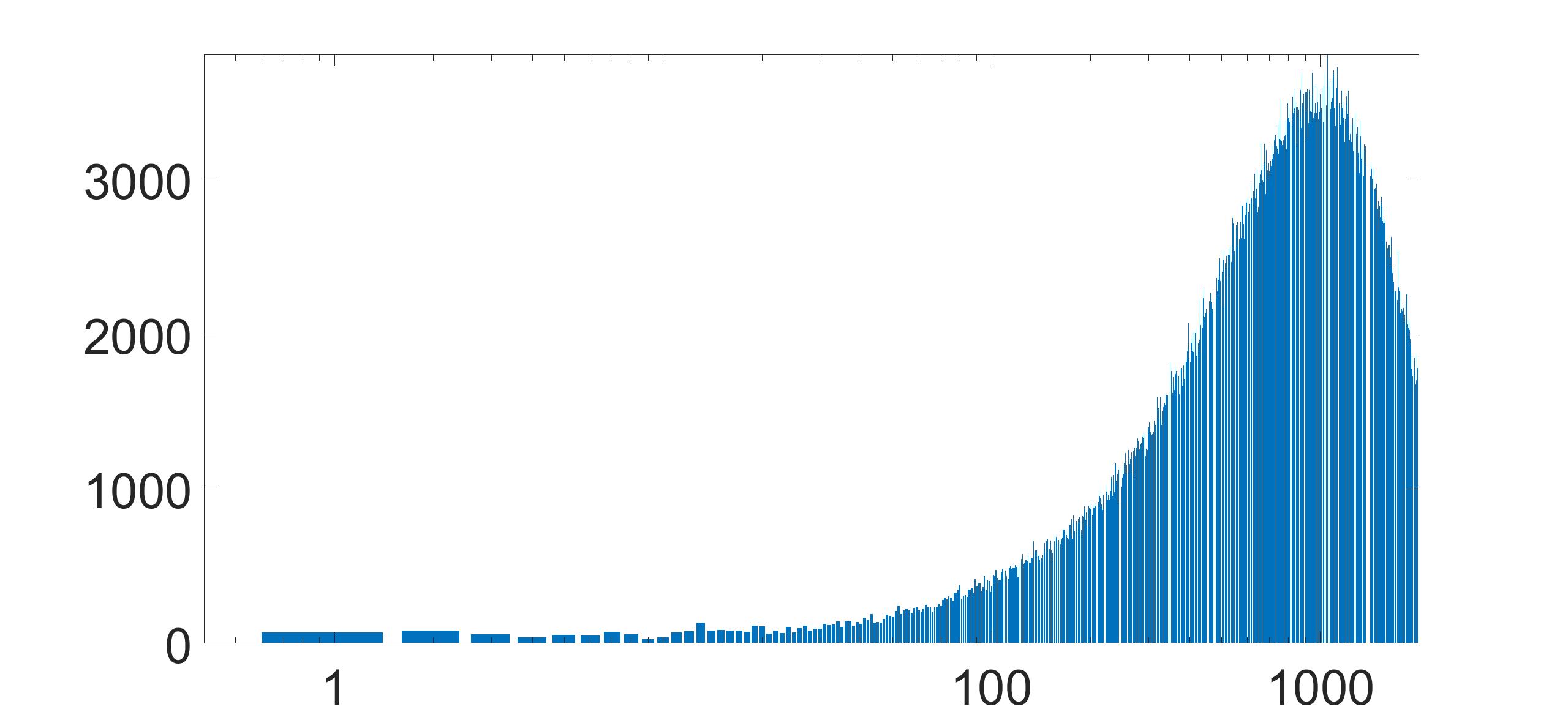, width=\linewidth}
      \caption{EASAL-Jacobian}
   \end{subfigure}
 \begin{subfigure}[b]{\wid}
      \epsfig{file = 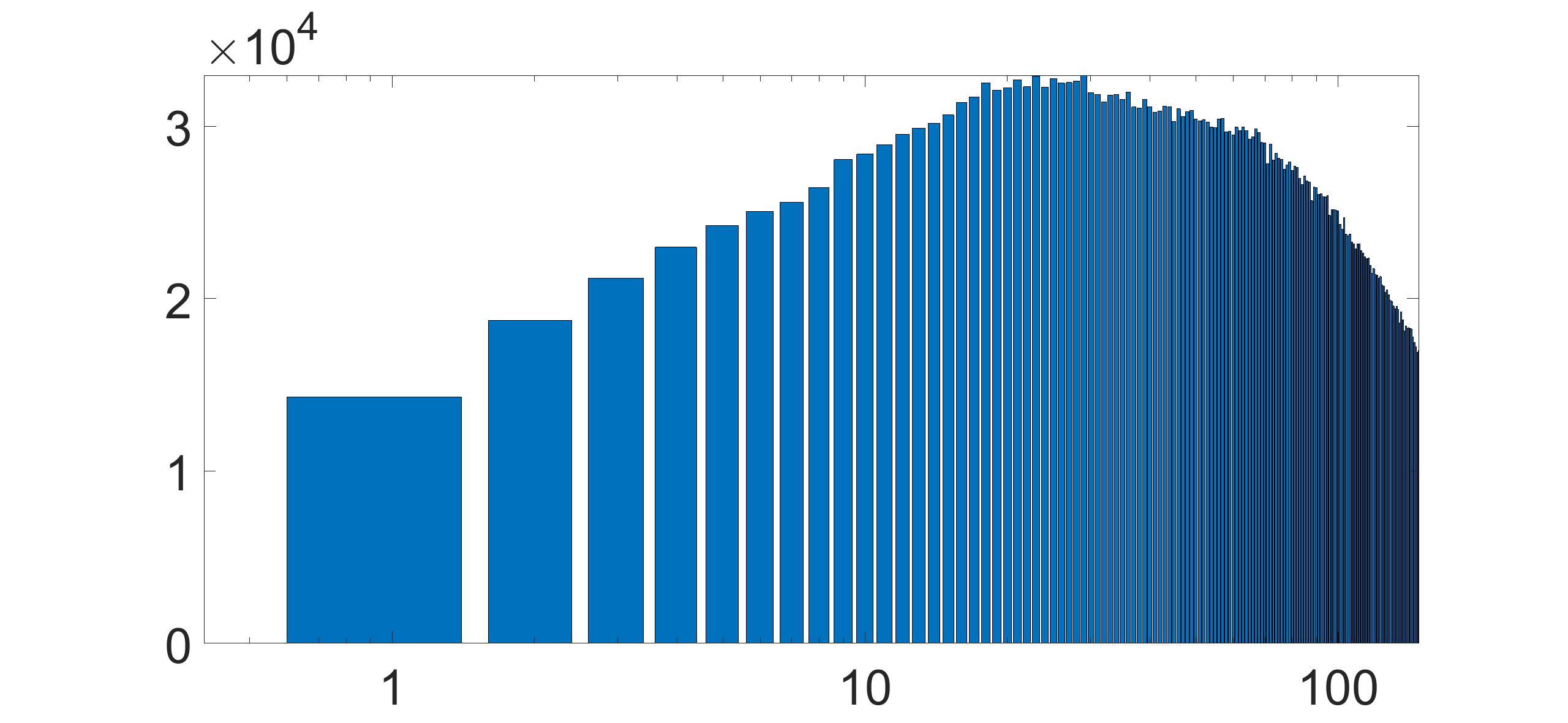, width=\linewidth}
      \caption{EASAL-1}
   \end{subfigure}
 \begin{subfigure}[b]{\wid}
      \epsfig{file = 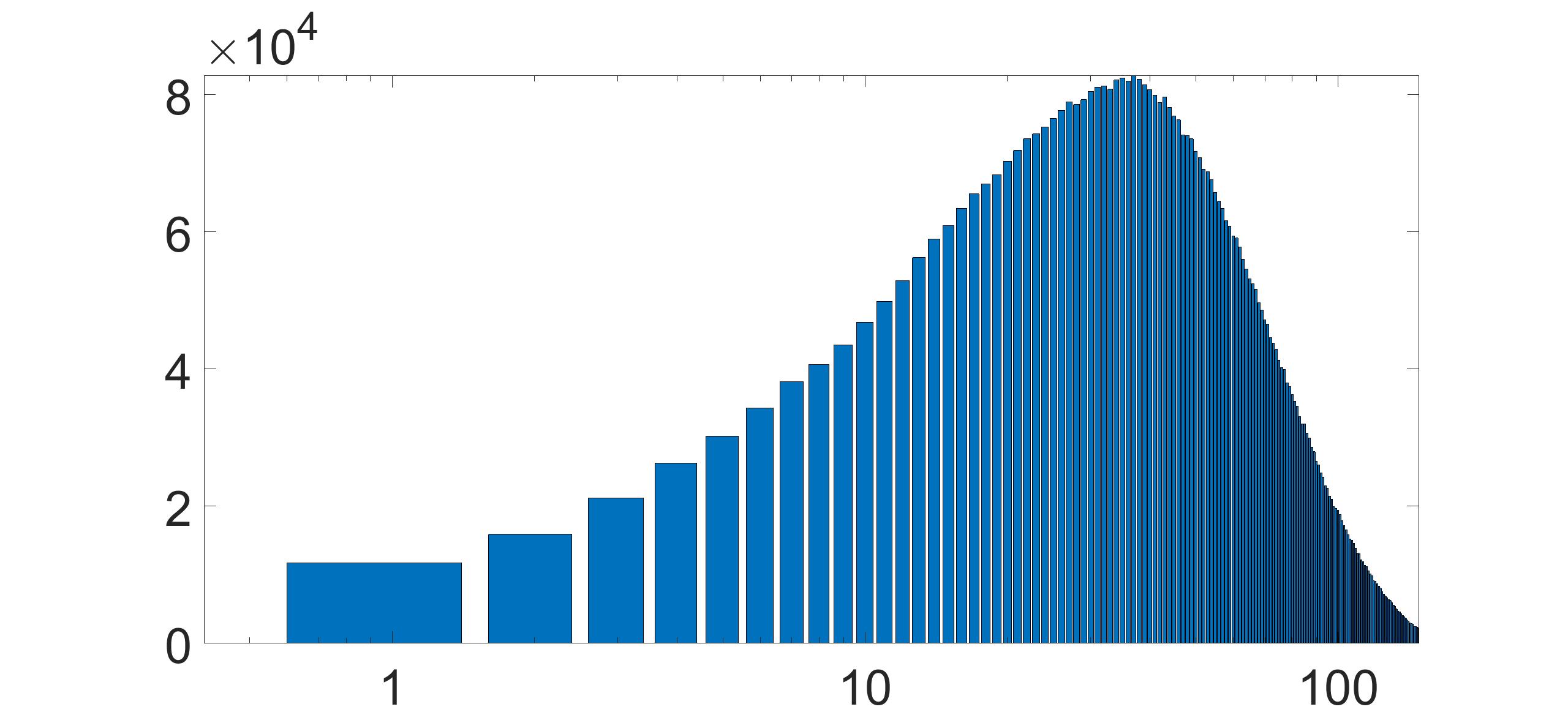, width=\linewidth}
      \caption{EASAL-2}
   \end{subfigure}
 \begin{subfigure}[b]{\wid}
      \epsfig{file = 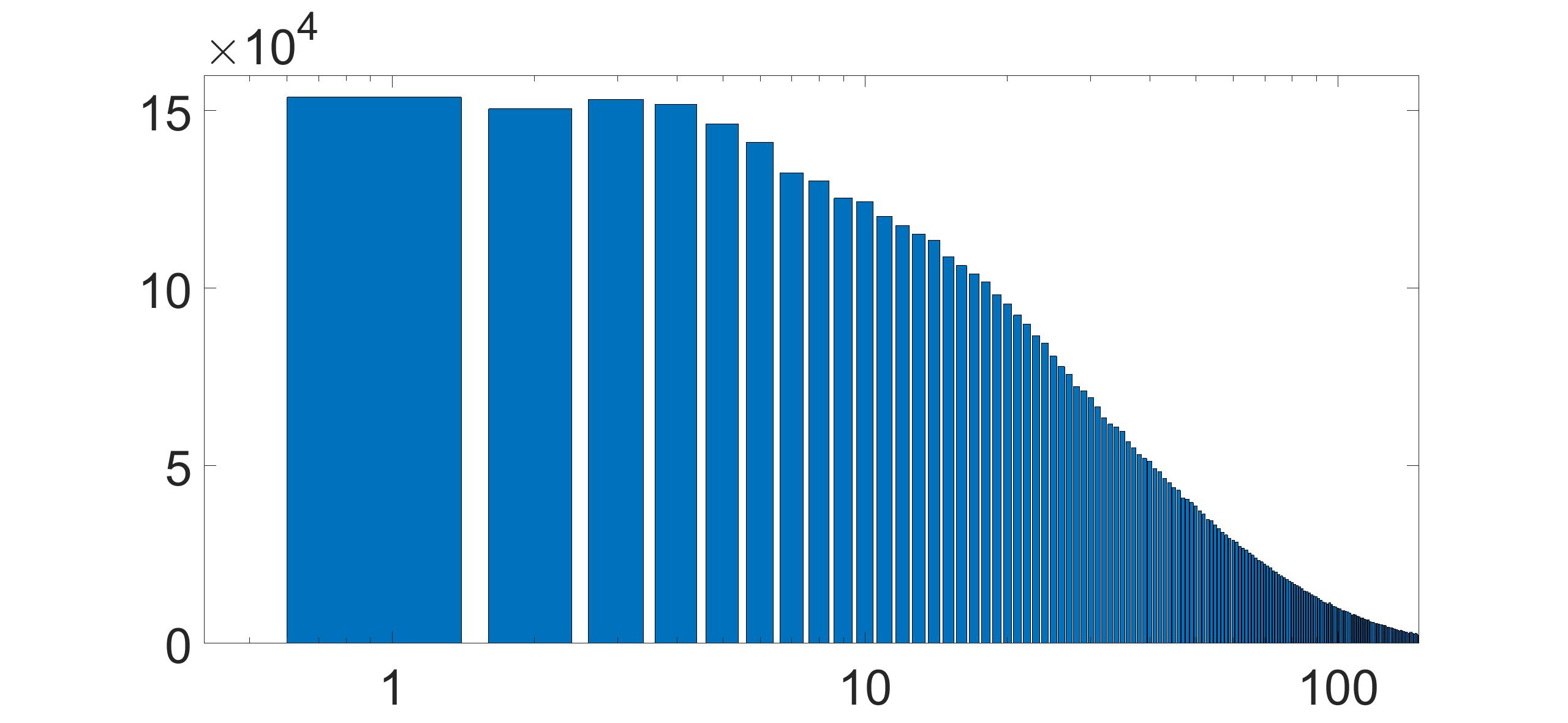, width=\linewidth}
      \caption{EASAL-3}
   \end{subfigure}
\caption{The horizontal axis shows the number of sample points $\nu$ that lie in $\varepsilon+1$-cubes,
and the vertical axis shows the number of $\varepsilon+1$-cubes having $\nu$ EASAL/MC-mapped points inside of them.
See Section \ref{sec:coverage}.
}
\label{fig:PointsVsRegions_Plus1}
\end{figure*}

The next step is trying to find an $\varepsilon$ for which we get denser
coverage from all the sampling algorithms. Figure
\ref{fig:PointsVsRegions_Plus1} plots the number of sample points $\nu$ (from
the different sampling algorithms) that lie in an $\varepsilon+1$-cube against
the number of $varepsilon+1$-cubes having $\nu$ EASAL/MC-mapped, sample points in them.  As
can be seen, expanding the size of the $\varepsilon$ cube results in much
denser coverage from all the sampling algorithms, with EASAL-3 still being
judicious with the number of samples per $\varepsilon$ cube.

\begin{table}[htpb]
\centering
\caption{Comparison of MC with EASAL variants for efficiency of uniform coverage,
with the two transmembrane helices shown in \figref{fig:atlas} as input. We show the
efficiency at both $\varepsilon$-coverage and $\varepsilon+1$-coverage.
See Section \ref{sec:coverage}.}
\label{tab:efficiency}
\begin{tabular}{|l|c|c|c|c|}\hline
		Sampling Algorithm & \multicolumn{2}{|c|}{$\varepsilon$-coverage} &\multicolumn{2}{c|}{$\varepsilon+1$-coverage}\\
		\cline{2-5}
		&$\mu_1$ & $\mu_2$ & $\mu_1$ & $\mu_2$\\\hline
		MC & 0.0002& 0 & 0.00001& 0\\\hline
		EASAL-1 &0.04& 2.81&0.008&0.14\\\hline
		EASAL-2 & 0.06& 9.87&0.002&0.29\\\hline
		EASAL-3 & 0.01& 25.1&0.02&5.1\\\hline
		EASAL-Jacobian & 0.002& 0.001&0.00001&0.00007\\\hline
\end{tabular}
\end{table}

Next, we use the two efficiency measures $\mu_1$ and $\mu_2$ described in
Section \ref{sec:keyMeasurements:UniformCoverage}, to compare the sampling
algorithms in terms of their efficiency at covering the uniform grid. Table
\ref{tab:efficiency} summarizes the results for coverage of $\varepsilon$ and
$\varepsilon+1$ cubes. EASAL variants have have better coverage efficiency in
terms of both measures. 

\subsection{Accuracy of Weighted Coverage}
\label{sec:LowerDim}
\begin{figure*} [htbp]
\def\wid{.25\textwidth}
\centering
	\begin{subfigure}[b]{\wid}
      \epsfig{file = 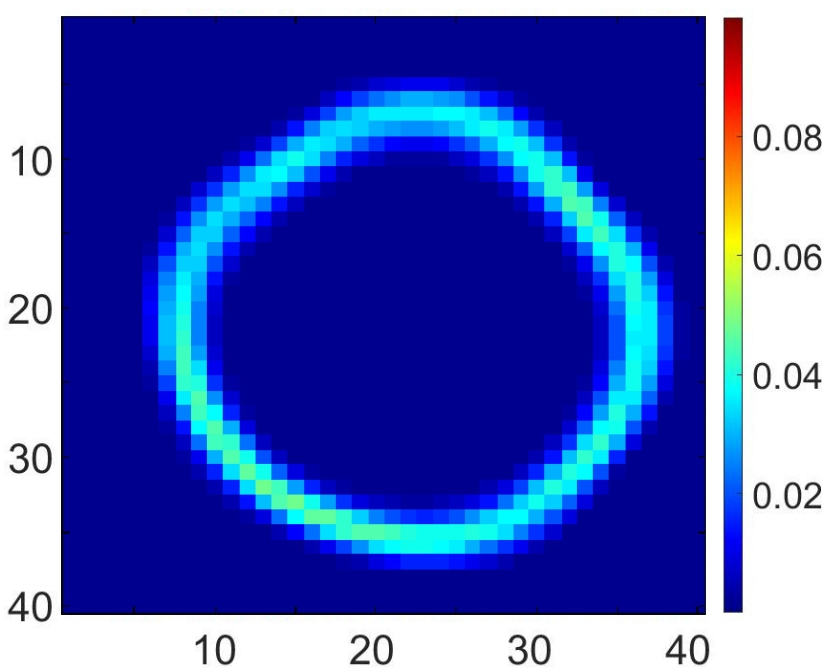, width=\linewidth}
      \caption{Grid}
   \end{subfigure}
 \begin{subfigure}[b]{\wid}
      \epsfig{file = 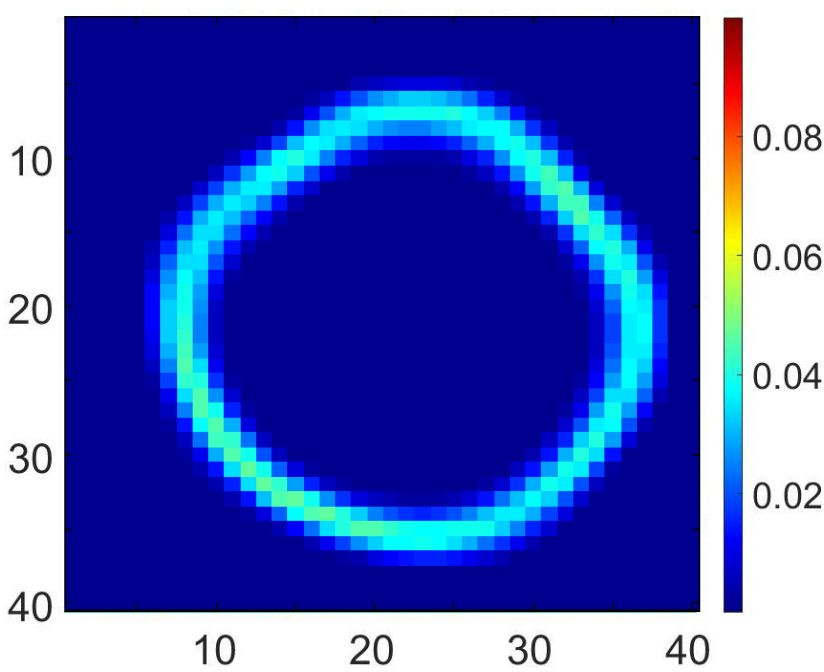, width=\linewidth}
      \caption{MC}
   \end{subfigure}
 \begin{subfigure}[b]{\wid}
      \epsfig{file = 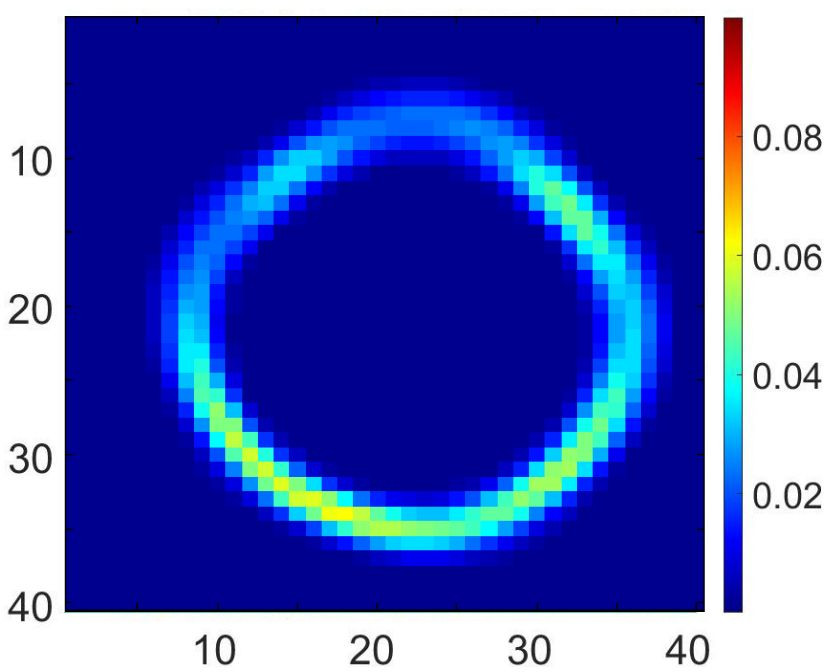, width=\linewidth}
      \caption{MultiGrid}
   \end{subfigure}
 \begin{subfigure}[b]{\wid}
      \epsfig{file = 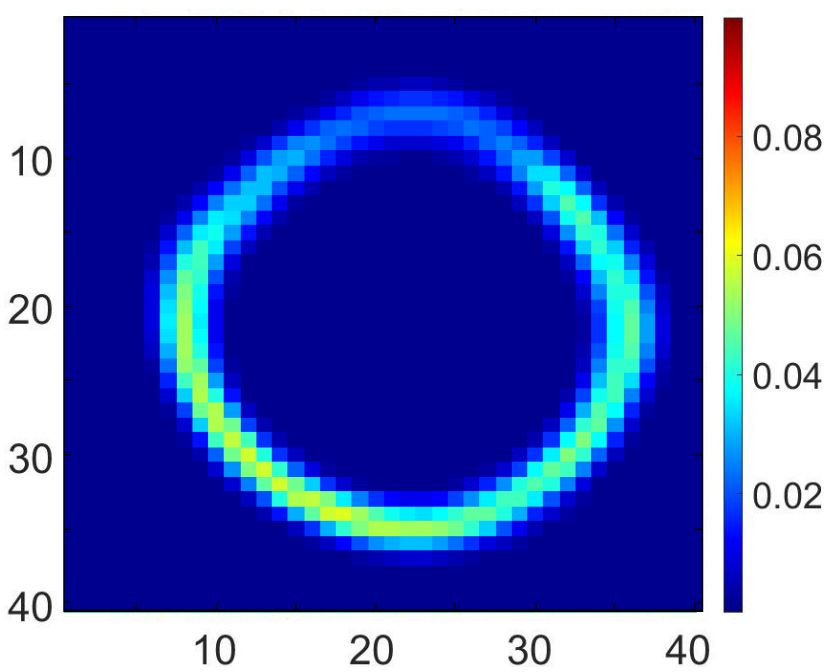, width=\linewidth}
      \caption{EASAL-Jacobian}
   \end{subfigure}
\begin{subfigure}[b]{\wid}
      \epsfig{file = 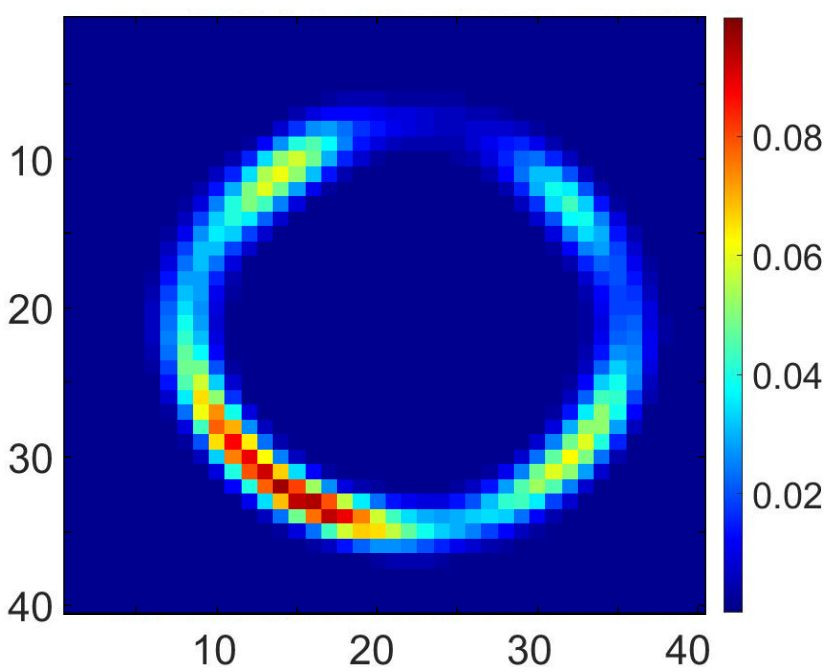, width=\linewidth}
      \caption{EASAL1}
   \end{subfigure}
 \begin{subfigure}[b]{\wid}
      \epsfig{file = 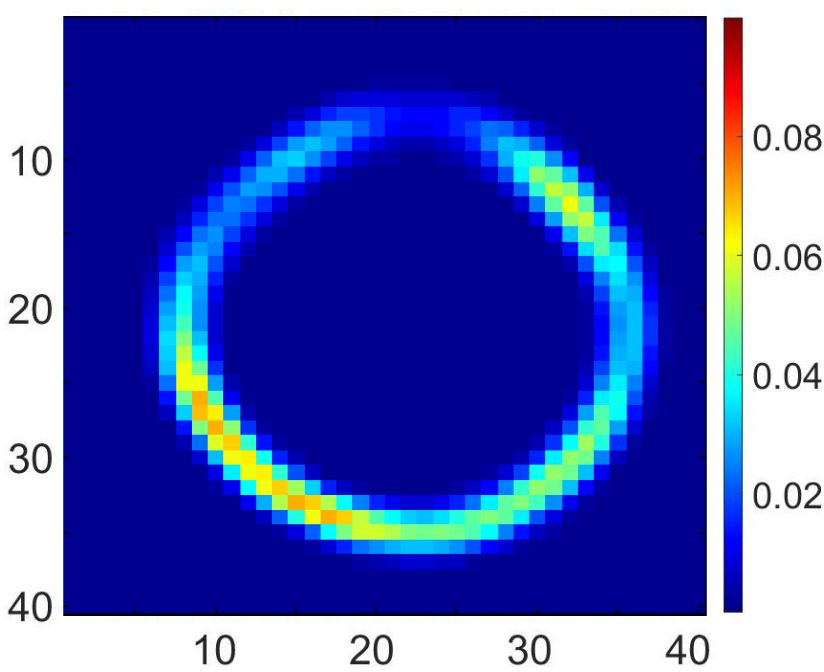, width=\linewidth}
      \caption{EASAL2}
   \end{subfigure}
 \begin{subfigure}[b]{\wid}
      \epsfig{file = 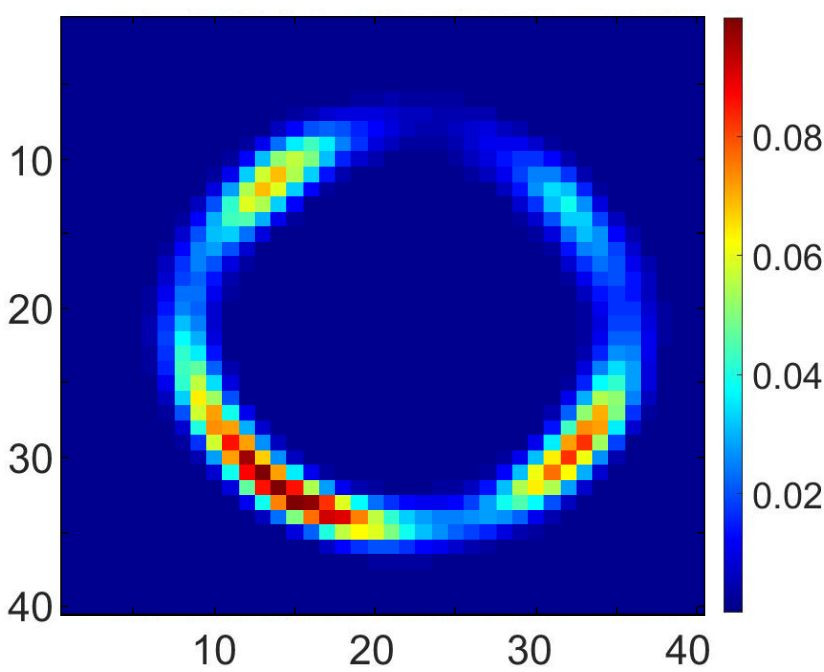, width=\linewidth}
      \caption{EASAL3}
   \end{subfigure}
\caption{ 
Projection in $R^2$ of a configuration space for the example helices described
		in Section \ref{sec:expSetup}, as sampled by different methods. The
		projection is on the $xy$ coordinates of the centroid of the second
		helix, with the centroid of the first helix fixed in the center. The
		color scale on the right of each figure corresponds to the number of
		sampled points in a $\varepsilon$-sized cube centered around the grid
		point $(x, y)$. See Section \ref{sec:LowerDim}.}
\label{fig:coverage_projection}
\end{figure*}

The objective of this experiment is to compare the performance of the sampling
methods at covering low energy regions. Figure \ref{fig:coverage_projection}
shows a 2-dimensional projection of the Cartesian configuration space of the
input transmembrane helices described in the experimental setup as sampled by
different methods. The projection is on the $xy$ coordinates of the centroid
of the second helix with the centroid of the first helix fixed at the origin.
There are no samples on the outer rim since since we have set the inter helical
angle to be less than $30^\circ$. There are also no samples in the center of
the plots because the configurations in this area have steric collisions
due to the helix mass center of the molecules being close to each other.

The sampling algorithm that covers lower energy region better, should have a
distribution of samples that resembles the MultiGrid figure closely. Notice
that EASAL-Jacobian and EASAL-2 approximate MultiGrid better than MC.

\begin{figure} [htbp]
\def\wid{.25\textwidth}
\centering
 \begin{subfigure}[b]{\wid}
		 \includegraphics[width=\linewidth]{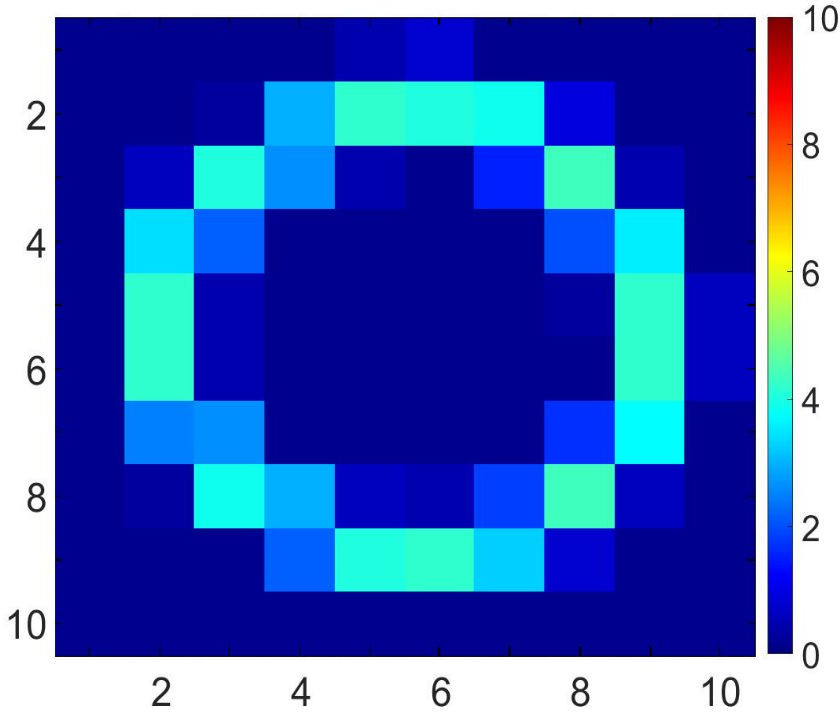}
      \caption{Grid \%}
   \end{subfigure}
 \begin{subfigure}[b]{\wid}
      \epsfig{file = 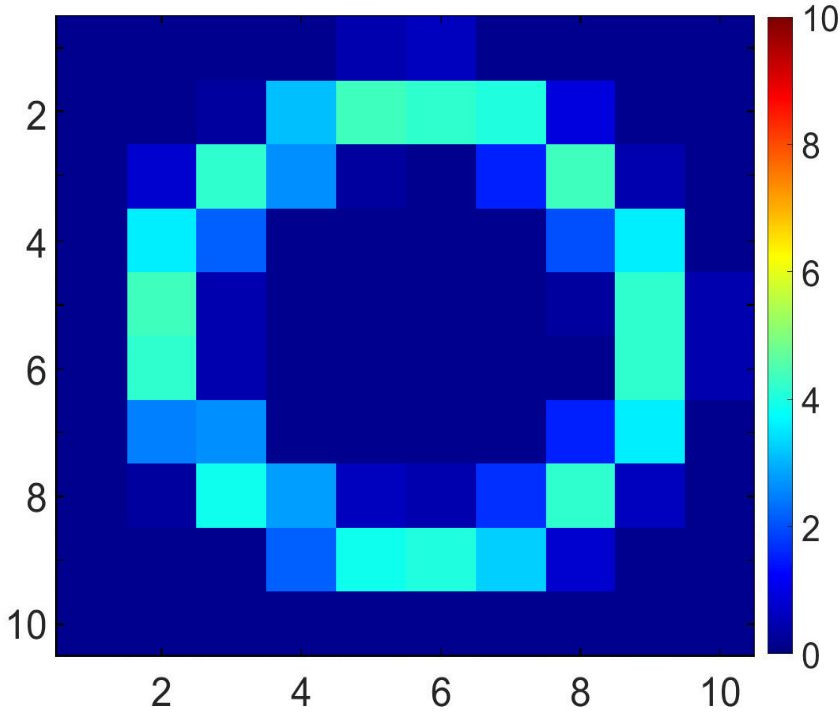, width=\linewidth}
      \caption{MC \%}
   \end{subfigure}
\begin{subfigure}[b]{\wid}
      \epsfig{file = 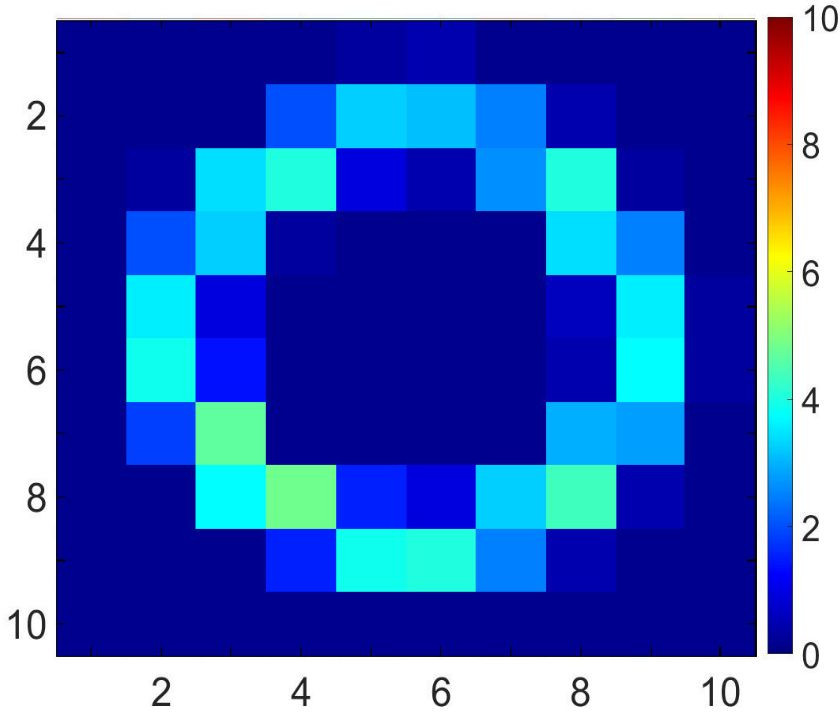, width=\linewidth}
      \caption{MultiGrid \%}
   \end{subfigure}
 \begin{subfigure}[b]{\wid}
      \epsfig{file = 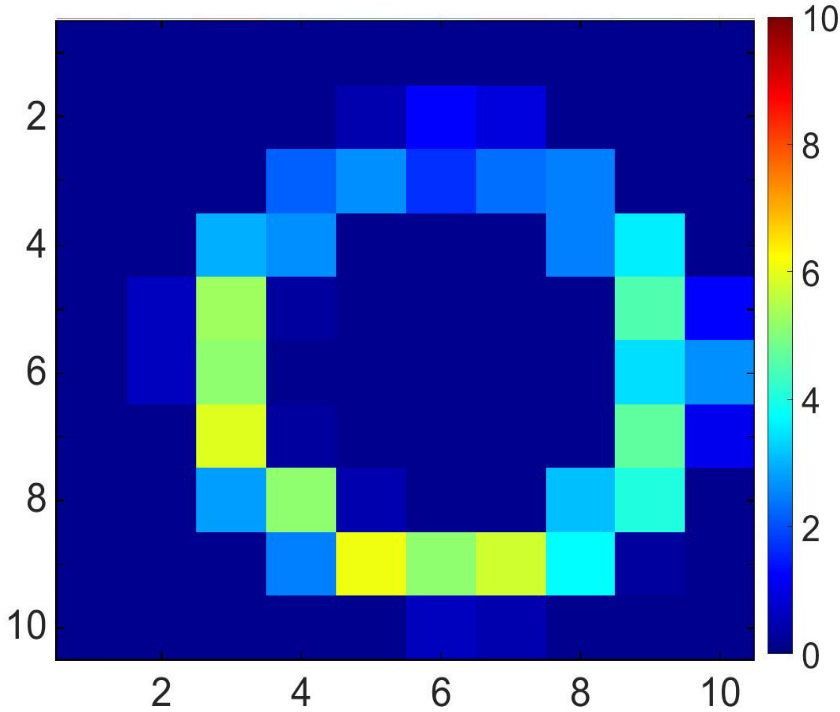, width=\linewidth}
      \caption{EASAL\_Jacobian \%}
   \end{subfigure}
 \begin{subfigure}[b]{\wid}
      \epsfig{file = 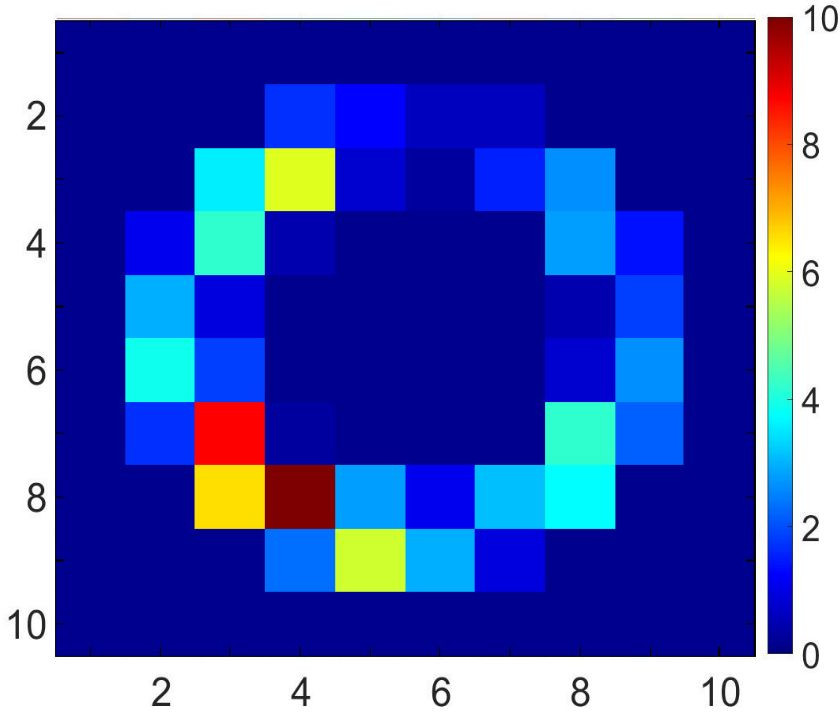, width=\linewidth}
      \caption{EASAL1 \%}
   \end{subfigure}
 \begin{subfigure}[b]{\wid}
      \epsfig{file = 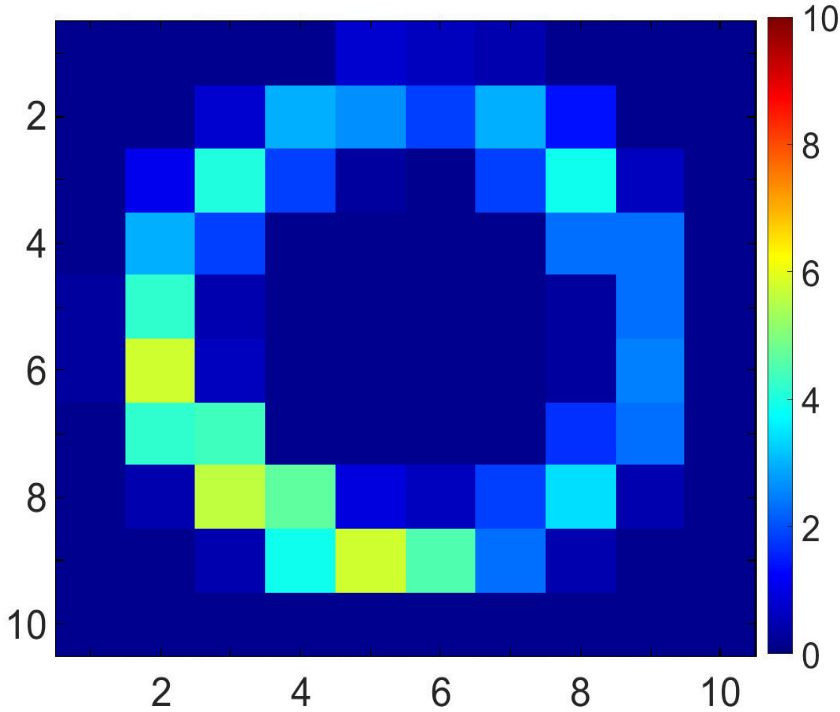, width=\linewidth}
      \caption{EASAL2 \%}
   \end{subfigure}
 \begin{subfigure}[b]{\wid}
      \epsfig{file = 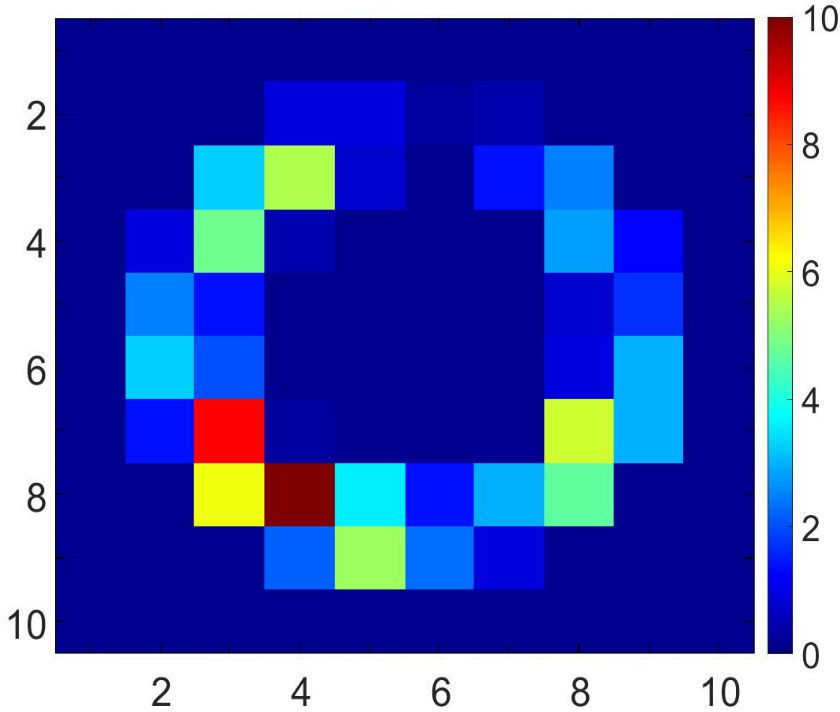, width=\linewidth}
      \caption{EASAL3 \%}
   \end{subfigure}
\caption{
Projection in $R^2$ of a configuration space for the example helices described
		in Section \ref{sec:expSetup}, as sampled by different methods. The
		projection is on the $xy$ coordinates of the centroid of the second
		helix, with the centroid of the first helix fixed in the center. We use a
		coarse $xy$ grid as described in the text. The
		color scale on the right of each figure corresponds to the
		percentage of samples in the coarse $xy$ region. 
		The percentage of samples is computed by summing all samples of subregions with a specific $x,y$
and dividing it by the total number of samples. See Section \ref{sec:LowerDim}.}
\label{fig:percentage_projection}
\end{figure}

In the next plot, we still take the $xy$ projection as before, but on a coarser
grid with 100 grid cubes (10 by 10) instead of the original 1600 grid cubes (40
by 40 used in Figure \ref{fig:coverage_projection}). Figure
\ref{fig:percentage_projection} plots the percentage of samples in an
$xy$-cube.  The percentage of samples is computed by summing all samples of
subregions with a specific $x,y$ and dividing it by the total number of
samples.

Since MultiGrid densely samples lower energy regions, the portions of the
MultiGrid plot with darker spots are the lower energy regions. As can be seen,
the EASAL variants sample these lower energy regions much denser than
MultiGrid. In contrast, MC and Grid, which are similar to each other, sample
these areas sparsely.

\begin{figure*} [htbp]
\def\wid{.25\textwidth}
\centering
 \begin{subfigure}[b]{\wid}
      \epsfig{file = 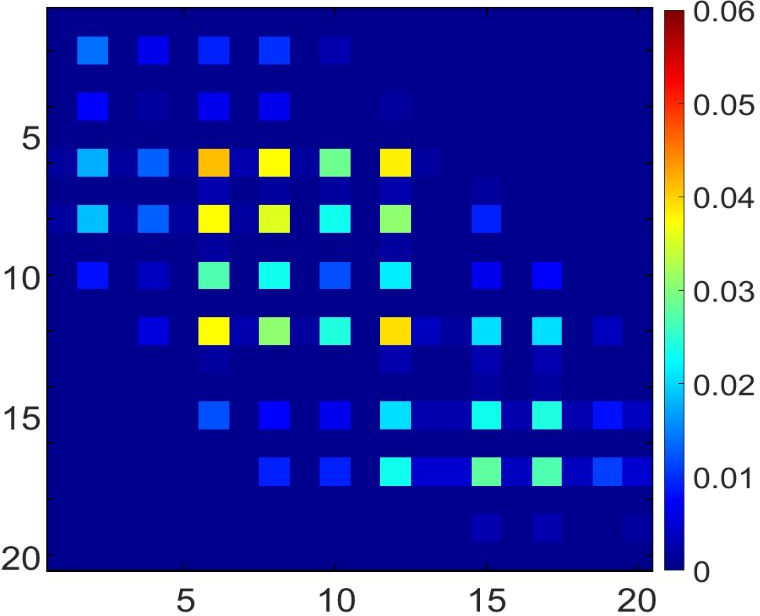, width=\linewidth}
      \caption{Atlas MC}
   \end{subfigure}
\begin{subfigure}[b]{\wid}
      \epsfig{file = 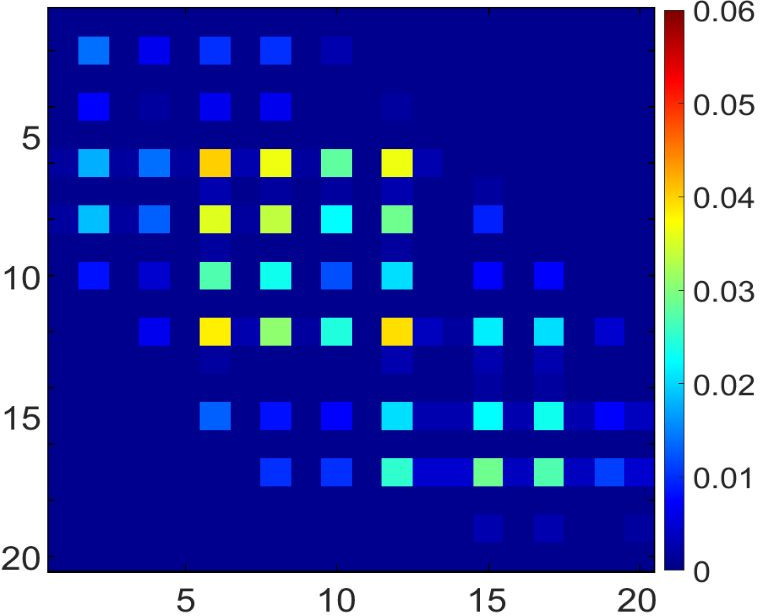, width=\linewidth}
      \caption{Atlas Grid}
   \end{subfigure}
\begin{subfigure}[b]{\wid}
      \epsfig{file = 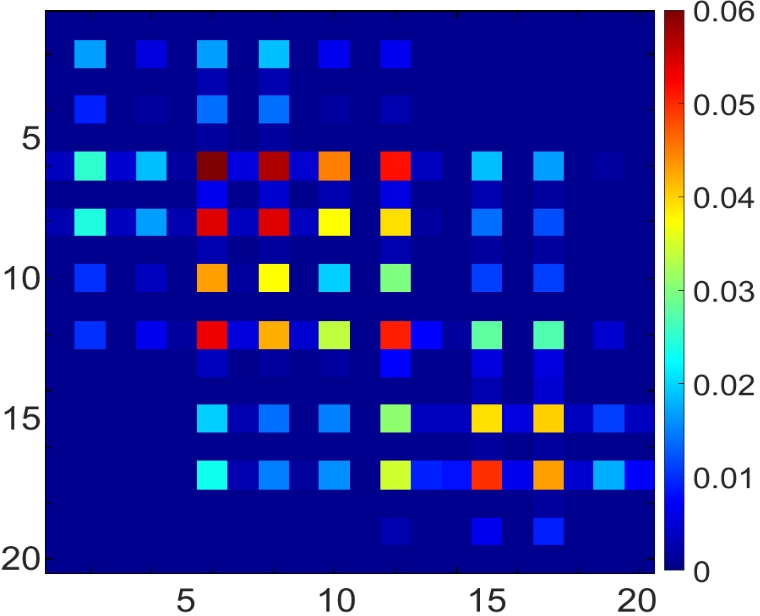, width=\linewidth}
      \caption{Atlas MultiGrid}
   \end{subfigure}
 \begin{subfigure}[b]{\wid}
      \epsfig{file = 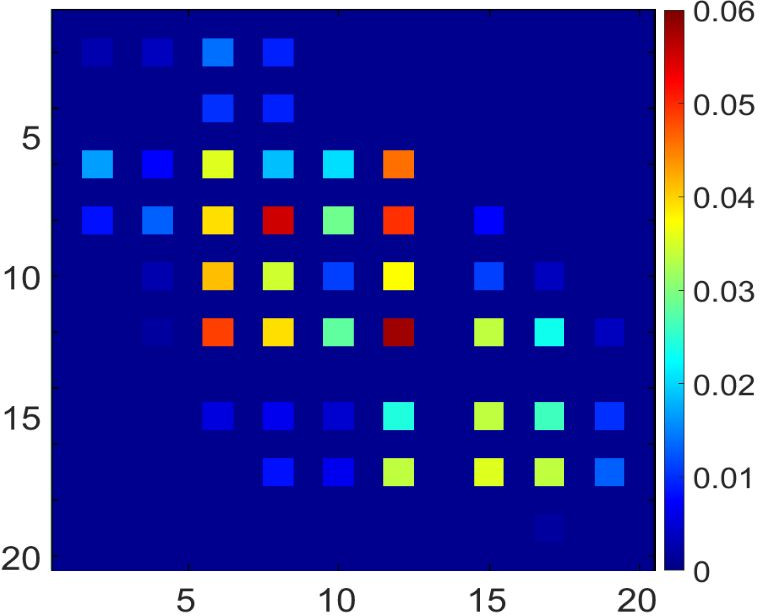, width=\linewidth}
      \caption{EASAL-Jacobian}
   \end{subfigure}
 \begin{subfigure}[b]{\wid}
      \epsfig{file = 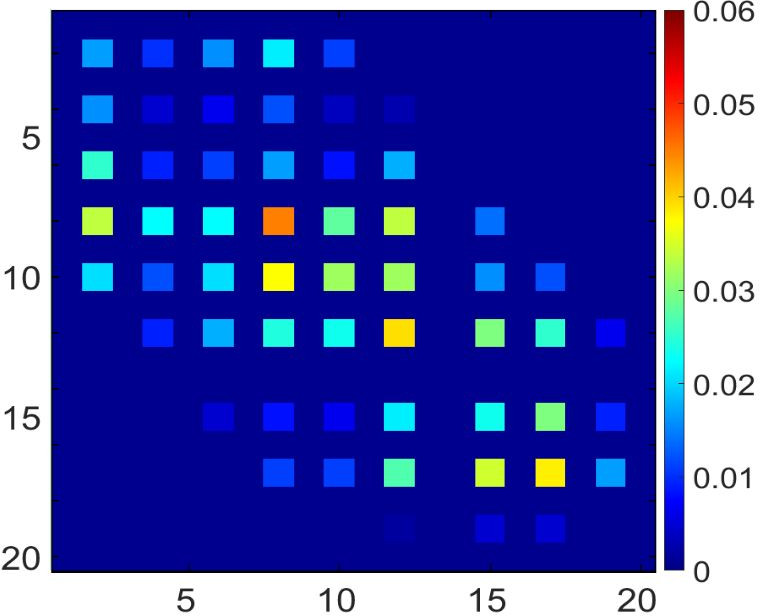, width=\linewidth}
      \caption{EASAL-1}
   \end{subfigure}
 \begin{subfigure}[b]{\wid}
      \epsfig{file = 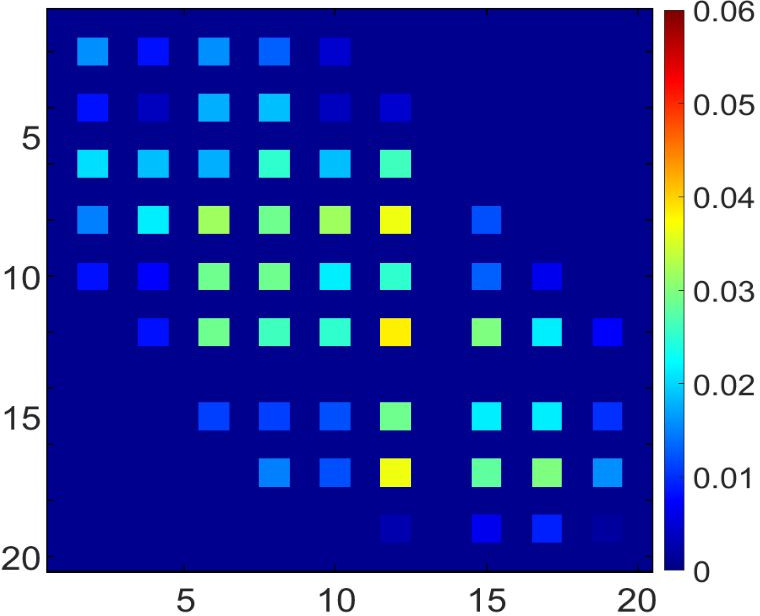, width=\linewidth}
      \caption{EASAL-2}
   \end{subfigure}
 \begin{subfigure}[b]{\wid}
      \epsfig{file = 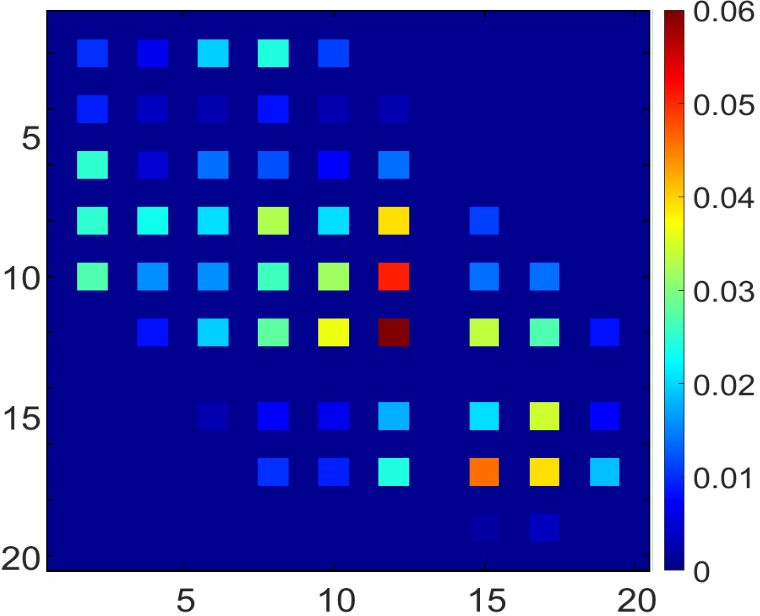, width=\linewidth}
      \caption{EASAL-3}
   \end{subfigure}
\caption{The atom indices of the two helices are shown as the rows and columns of a matrix.
		The $(row_1, column_1)$ entry represents a 5D active constraint region, where
		the active constraint is between the atom number $row_1$ on the first molecule
		and atom number $column_1$ on the second molecule. 
		The color code gives the distribution ratio. The plot of atlas MultiGrid
		has higher distribution ratios for active constraint regions with where
		there are more lower energy configurations. Note that EASAL variants resemble
		distribution ratios of MultiGrid more than grid and MC distributions.
		See Section \ref{sec:distribution}.}
\label{fig:distRatio}
\end{figure*}

Figure \ref{fig:distRatio} plots the distribution of samples across different
active constraint regions or atlas nodes.  The atom indices of the two helices
are shown as the rows and columns of a matrix.  The $(row_1, column_1)$ entry
represents a 5D active constraint region, where the active constraint is
between the atom number $row_1$ on the first molecule and atom number
$column_1$ on the second molecule.  The color code gives the distribution
ratio. The plot of atlas MultiGrid has higher distribution ratios for active
constraint regions with where there are more lower energy configurations. Note
that EASAL variants resemble distribution ratios of MultiGrid more than grid
and MC distributions.

\subsection{Localized Sampling of Individual Active Constraint Regions (macrostates)}
\label{sec:distribution}
The goal of this experiment is to demonstrate EASAL's ability to restrict
sampling to specified macrostates. 

\begin{table}[htpb]
\centering
\caption{Comparison of MC with EASAL variants in terms of ratio percentages.
See Section \ref{sec:distribution}.}
\label{tab:ratioPercentage}
\resizebox{\textwidth}{!}{%
\begin{tabular}{lccccc}\hline
	Sampling method & EASAL-1 &EASAL-2& EASAL-3& EASAL-Jacobian& MC \\\hline
	Ratio percentage & 3.56\%& 5.17\%& 2.97\% &3.45\% &1.29\% \\\hline
\end{tabular}
}
\end{table}

\begin{figure}[htpb]
	\centering
	\includegraphics[width=\columnwidth]{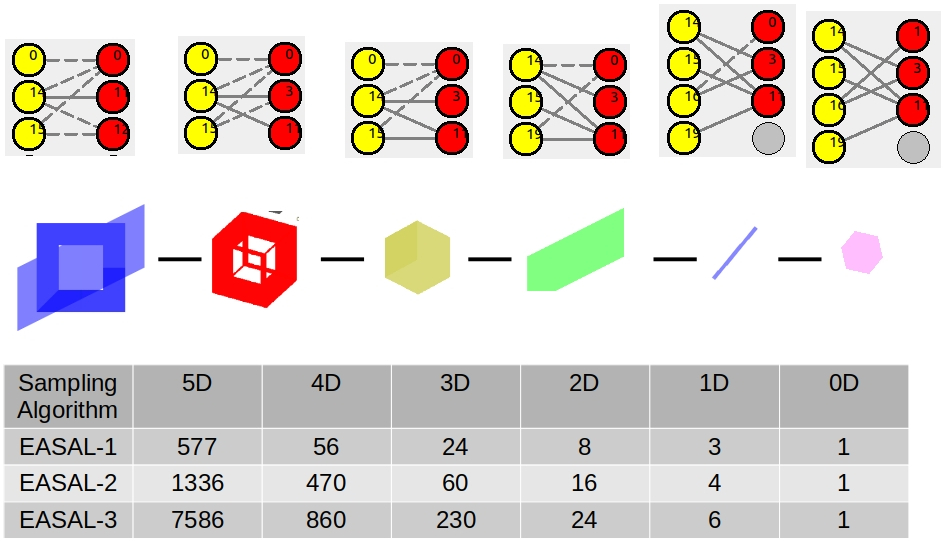}
	\caption{A path in a the EASAL atlas from a 5D to a 0D node, showing the active constraint 
	graphs of each node and the number of samples required to sample each region with different
	variants of EASAL. See Section \ref{sec:distribution}.}
	\label{fig:flexibility}
\end{figure}
Table \ref{tab:ratioPercentage} summarizes the ratio percentages of MC and EASAL variants
for sampling the two transmembrane helices described in the experimental setup. 
The ratio percentage gives a measure of the percentage of samples required to directly
sample a 3D region as compared to trying to sampling the entire energy landscape.
For instance, EASAL-1 has ratio percentage of 3.56\%, this means that EASAL-1 would
need only 3.56\% of samples to sample a 3D active constraint region or macrostate
directly, as compared to compared to sampling portion of the configuration space
starting with a 5D active constraint region and reaching the given 3D region.
EASAL's ability to localize sampling reduces the number of samples required to sample
particular macrostates. 

This is further illustrated in Figure \ref{fig:flexibility} with a concrete
example. We show a path in the EASAL atlas from a 5D to a 0D active constraint
region along with the number of samples required to sample each of them using
different EASAL variants. The table in Figure \ref{fig:flexibility} shows how
being able to localize sampling to a particular macrostate can reduce the
number of samples required, especially when sampling the lower dimensional
regions. For instance, if we are interested in only sampling the 3D region
shown in the figure, the EASAL variants would need 24, 60, and 230 samples.
However, in the absence of this ability, we would have to sample the entire
configuration space to sample just this region, which is the case with MC.

However, as shown in the paper \cite{maineasal}, EASAL gives the best accuracy
when starting sampling from 5D active constraint regions. When the starting
dimension of sampling goes down, EASAL's accuracy in terms of the number of 0D
regions discovered goes down. Despite this, EASAL is still able to discover a
respectable number of 0D active constraint regions even with a very coarse
sampling step size and starting from the 3D region shown in Table
\ref{tab:flexibility}.

\begin{table}[htpb]
\centering
\caption{The number of 0D regions discovered with EASAL-1 sampling,
when starting sampling from the 3D region shown in Figure 
\ref{fig:flexibility} with varying step sizes.}
\label{tab:flexibility}
\begin{tabular}{cc}\hline
	Step size & 0D regions discovered \\\hline
	0.4 & 144\\\hline
	0.8 & 122\\\hline
	1.2 & 82\\\hline
	1.6 & 72\\\hline
	2 & 60\\\hline
\end{tabular}
\end{table}

\section{Discussion}
\label{sec:discussion}
The uniform grid described earlier, can be sampled by brute force to get
good coverage. However, this uses a large number of samples and
over 90\% of these are typically discarded since they are high energy
configurations (no active constraints). This method also finds it difficult to
sample low energy (more than 1 active constraint) regions.

In the case of MC, the simulation has a tendency to sample around a localized
sub-set of low energy configurations that are located close to each other. To ensure
coverage, one needs to compensate for the lack of ergodicity by computationally
intensive use of a large number of samples.

EASAL has flexibility in sampling distributions, but in all cases guarantees
reasonable coverage of low energy regions even if they have low effective
dimension. It tends to over-sample these lower energy regions (i.e, regions
where more pairs of atoms are in their Leonard-Jones well). Hence EASAL can
help MC to find high energy barriers and force the MC simulations to move to
the locations of low energy basins. Macrostates with lower energy, which MC
misses due to broken ergodicity, can be sampled using EASAL and used as a seed
for MC in order to generate the free energy profile for the whole configuration
space. 

More precisely,  we expect EASAL to be used to help evaluate/improve MC with
the following path.
\begin{enumerate}
\item Run a coarse-grained version of MC with the usual energy function
and a relatively small number of samples

\item Run EASAL with constraints extracted from the MC energy function, verify
that the space sampled by EASAL encompasses the relevant space. 

\item List description of macrostates that MC missed, and give an
estimate of the volume of the regions.

\item Compute energy on these regions, and if low, then run MC seeded on these
(lower volume) regions (from which larger volume regions can be
reached), in order to generate the free energy profile for the
whole configuration space.

\item Compare new trajectories with old MC trajectories.
\end{enumerate}

\section{Conclusion}
\label{sec:conclusion}
We compare the sampling characteristics of the EASAL methodology to the
traditional MonteCarlo sampling, using custom-designed measurements, for
sampling the the assembly landscape of 2 transmembrane helices, with
short-range pair-potentials.  We demonstrate that variants of EASAL provide
good coverage of narrow regions of low potential energy and low effective
dimension with much fewer samples and computational resources than MC.

We provide promising avenues for combining the complementary advantages of the
two methods, which can significantly improve the current state of the art of
free energy and other integral computations, specifically for small assemblies.
EASAL is tailored for such assemblies and can be used to improve accuracy and
ergodicity guarantees for Monte Carlo trajectories. It can also help explain
the behavior of MC trajectories by using EASAL to infer geometric and
topological features of the configuration space.

EASAL can help MC find high energy barriers and force the simulations to
move to the locations of low energy basins. Finally, lower energy regions
located by EASAL can be used as seeds for MC trajectories, to speed up
traversal of the entire configuration space.

\bibliographystyle{unsrt}
\bibliography{easal,pr,stickysphere,nigms,jorg,Dmay04}

\end{document}